\documentclass[
 reprint,
 showpacs, preprintnumbers,
 amsmath, amssymb,
 aps,
 prl,
 floatfix,
 superscriptaddress
]{revtex4-2}

\usepackage{amsmath}
\usepackage{graphicx}
\usepackage{dcolumn}
\usepackage{bm}
\usepackage{bbm}

\usepackage[utf8]{inputenc}
\usepackage[T1]{fontenc}
\usepackage{mathptmx}
\usepackage{textgreek}
\usepackage{upgreek}
\usepackage{float}
\usepackage{hyperref}

\usepackage{color}

\usepackage{natbib}

\makeatletter
\def\maketag@@@#1{\hbox{\m@th\normalfont\normalsize#1}}
\makeatother

\newcommand{\beginsupplement}{%
        \setcounter{table}{0}
        \renewcommand{\thetable}{S\arabic{table}}%
        \setcounter{figure}{0}
        \renewcommand{\thefigure}{S\arabic{figure}}%
				\renewcommand{\theequation}{S.\arabic{equation}}
     }

\begin{document}

	\preprint{}

\title[JJ Diode]{A Josephson junction supercurrent diode} 

\affiliation{Institut f\"ur Experimentelle und Angewandte Physik, University of Regensburg, 93040 Regensburg, Germany}
\author{C.~Baumgartner$^\dagger$}
\author{L.~Fuchs$^\dagger$}
\affiliation{Institut f\"ur Experimentelle und Angewandte Physik, University of Regensburg, 93040 Regensburg, Germany}
\author{A.~Costa}
\affiliation{Institut f\"ur Theoretische Physik, University of Regensburg, 93040 Regensburg, Germany}
\author{S.~Reinhardt}
\affiliation{Institut f\"ur Experimentelle und Angewandte Physik, University of Regensburg, 93040 Regensburg, Germany}

\author{S.~Gronin}
\author{G.~C.~Gardner}
\affiliation{Microsoft Quantum Purdue, Purdue University, West Lafayette, Indiana 47907 USA}
\affiliation{Birck Nanotechnology Center, Purdue University, West Lafayette, Indiana 47907 USA}
\author{T.~Lindemann}
\affiliation{Department of Physics and Astronomy, Purdue University, West Lafayette, Indiana 47907 USA}
\affiliation{Birck Nanotechnology Center, Purdue University, West Lafayette, Indiana 47907 USA}

\author{M.~J.~Manfra}
\affiliation{Birck Nanotechnology Center, Purdue University, West Lafayette, Indiana 47907 USA}
\affiliation{Microsoft Quantum Purdue, Purdue University, West Lafayette, Indiana 47907 USA}
\affiliation{Department of Physics and Astronomy, Purdue University, West Lafayette, Indiana 47907 USA}
\affiliation{School of Materials Engineering, Purdue University, West Lafayette, Indiana 47907 USA}
\affiliation{School of Electrical and Computer Engineering, Purdue University, West Lafayette, Indiana 47907 USA}

\author{P.~E.~Faria~Junior}
\author{D.~Kochan}
\author{J.~Fabian}
\affiliation{Institut f\"ur Theoretische Physik, University of Regensburg, 93040 Regensburg, Germany}

\author{N.~Paradiso}\email{nicola.paradiso@physik.uni-regensburg.de}
\author{C.~Strunk}
\affiliation{Institut f\"ur Experimentelle und Angewandte Physik, University of Regensburg, 93040 Regensburg, Germany}
%


\begin{abstract}	
    
Transport is called nonreciprocal when not only the sign, but also the absolute value of the current, depends on the polarity of the applied voltage.  It requires simultaneously broken inversion and time-reversal symmetries, e.g., by the interplay of spin-orbit coupling and magnetic field. So far, observation of nonreciprocity was always tied to resistivity, and dissipationless nonreciprocal circuit elements were elusive. Here, we engineer fully superconducting nonreciprocal devices based on highly-transparent Josephson junctions fabricated on InAs quantum wells. We demonstrate supercurrent rectification far below the transition temperature. By measuring Josephson inductance, we can link nonreciprocal supercurrent to the asymmetry of the current-phase relation, and directly derive the supercurrent magnetochiral anisotropy coefficient for the first time. A semi-quantitative model well explains the main features of our experimental data. Nonreciprocal Josephson junctions have the potential to become for superconducting circuits what $pn$-junctions are for traditional electronics, opening the way to novel nondissipative circuit elements.
\end{abstract}

\maketitle

The $pn$-junction~\cite{ScaffOhl1947,Shockley1949} is a key component of modern electronics, being the fundamental building block of, among other things, LEDs, current rectifiers, voltage-controlled oscillators, photosensors, and solar cells.
Its working principle is the rectification effect produced by the spatial asymmetry of the junction, i.e., the non-equivalence of the $p$- and $n$-leads. In \textit{homogeneous} devices, the resistance depends instead on the current direction only if both inversion and time-reversal symmetry are simultaneously broken. This can be obtained, for example, by applying an electric and a magnetic field perpendicular to each other.

In the linear response regime, the absence of both parity and time-reversal symmetry produces, by the Onsager relations~\cite{OnsagerPR31,Kubo57,Tokura2018,Hoshino2018}, longitudinal transport coefficients that depend on the polarity of the current. Rikken \textit{et al.}~\cite{RikkenPRL2001,RikkenPRL2005} extended this result to the nonlinear regime and to diffusive 2D conductors. They found that, if the  electric field $\vec{E}\parallel \hat{e}_z$ is directed out-of-plane, whereas the magnetic field $\vec{B}$ and current $\vec{I}$ are directed in-plane, the resistance can be written as
\begin{equation}
R=R_0[1+\gamma \hat{e}_z(\vec{B}\times\vec{I})].
\label{eq:rikken}
\end{equation}
In normal conductors, the magnetic field-dependent term is usually tiny---a consequence of the fact that the spin-orbit interaction~(SOI) is typically many orders of magnitude smaller than the Fermi energy. However, in recent years several experiments have demonstrated a robust nonreciprocal charge transport in the fluctuation regime of noncentrosymmetric superconductors~\cite{Wakatsuki2017,Itahashi2020}. Here, the energy scale, to which the SOI has to be compared to, is not the Fermi energy, but the superconducting gap. As a result, the resistance shows a sizable magnetochiral anisotropy coefficient $\gamma=\gamma_S$~\cite{Tokura2018,Hoshino2018,Wakatsuki2017}. The observation of nonreciprocal transport in superconductors was so far confined to a narrow temperature window near $T_\mathrm{c}$, as it relies on measuring DC resistance, which vanishes at low temperatures. On the other hand, in this regime the supercurrent response to an AC excitation is described by its superfluid stiffness, which can be detected via kinetic inductance measurements. A natural question is whether a magnetochiral~anisotropy exists for the superfluid, described by the analog of Eq.~\ref{eq:rikken}, 
\begin{equation}
L=L_0[1+\gamma_L \hat{e}_z(\vec{B}\times\vec{I})], 
\label{eq:gammaell}
\end{equation}
with the resistance substituted by the kinetic inductance $L$.  The magnetochiral anisotropy coefficient $\gamma_L$ constitutes a novel observable characterizing nonreciprocal supercurrents.

\begin{figure*}[tb]
\includegraphics[width=2\columnwidth]{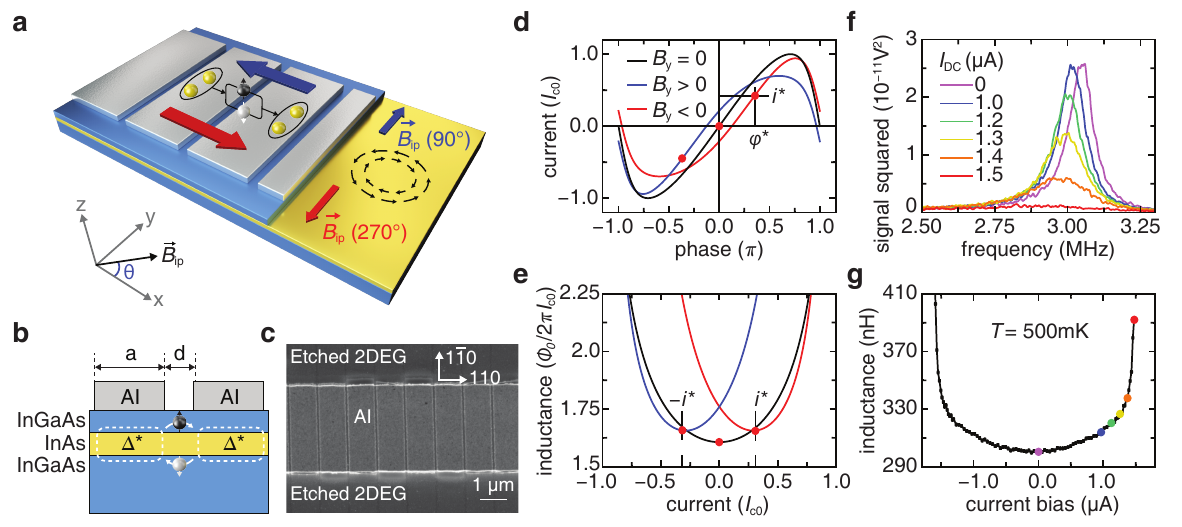}
\caption{\textbf{Josephson junction array and anomalous current--phase relation}. \textbf{a},~Sketch of Josephson junction array formed by a chain of Al~islands (grey) on top of an InAs~quantum well (yellow). Red and blue arrows denote the spontaneous supercurrents flowing at zero phase difference via spin-split  pairs of Andreev bound states (black loop). Their strength and direction depends on that of a transverse magnetic field $\vec{B}_\mathrm{ip}$, which competes with the Rashba spin-texture (counterpropagating circles of black arrows) in the quantum well. The actual array is made of 2250 islands of width $w=3.15$~\textmu m and length $a=1.0$~\textmu m. The island separation is $ d = 0.1$~\textmu m. \textbf{b},~Growth sequence for the heterostructure under study. \textbf{c},~Scanning electron micrograph of a part of the array, taken before the fabrication of a global top~gate. \textbf{d},~Illustrative current-phase relation (CPR) for a short-ballistic Josephson junction with high transparency $\tau=0.94$ and strong SOI in the absence (black) and presence of a transverse magnetic field $\vec{B}_{y} \parallel \hat{y}$ (red: $B_y>0$, blue: $B_y<0$). In this example, the effect of finite $\pm B_y$ is to reduce by a factor 0.8 the critical current, $I_c=0.8I_{c0}$, and to add a cosinusoidal term $\pm 0.2I_c\cos (\varphi)$ to the CPR's Fourier series. The red dots denote the inflection points ($i^\ast,\varphi^\ast$) of the CPR. \textbf{e},~Corresponding Josephson inductance as function of current. \textbf{f},~Resonance curves for the RLC~circuit, measured at 500~mK for different values of the current bias. \textbf{g},~Measured Josephson inductance $L(I)$ vs. current at $B=0$. Colored symbols correspond to the spectra in~\textbf{f}.
}
\label{fig:firstfig}
\end{figure*}

In this work, we have engineered nonreciprocal superflow in synthetic noncentrosymmetric superconductors, i.e., arrays of highly-transparent Josephson junctions~(JJs) based upon an InAs 2D electron gas~(2DEG) proximitized by epitaxial Al. Using a recently demonstrated resonator technique~\cite{baumgartner2020}, we measure the kinetic (or Josephson) inductance and investigate the effects of simultaneous inversion and time-reversal symmetry breaking on the current--phase~relation~(CPR). If the in-plane magnetic field has a component perpendicular to the current direction, a pronounced asymmetry appears in the current dependence of the Josephson inductance $L(I)$ that signals 
the magnetochiral anisotropy of the supercurrent, defined in Eq.~\ref{eq:gammaell}. This means that the supercurrent is different for opposite polarities of the phase difference, corresponding to a partial \textit{rectification} of supercurrent.
%

Our devices, sketched in Figs.~\ref{fig:firstfig}\textbf{a},\textbf{b} are fabricated starting from an InAs quantum well hosting a shallow 2DEG separated by a 10~nm thick InGaAs barrier from a 7~nm thick Al epitaxial film. The Al films induces in the 2DEG a superconducting gap $\Delta^{\ast}$ of about 130~\textmu eV by proximity effect. SNS Josephson junctions are obtained by patterning the Al film by electron-beam~lithography and selective etching. The etching process leaves an array of $N$ $3.15$~\textmu m-wide and $1.0$~\textmu m-long rectangular Al/2DEG islands separated by 3.15~\textmu m-wide and $0.1$~\textmu m-long areas with the Al film stripped off, which serve as weak links. A scanning electron micrograph of part of the array is shown in Fig.~\ref{fig:firstfig}\textbf{c}. The sample is embedded in a RLC~circuit that is integrated into the sample holder. The circuit allows us to measure both DC current-voltage ($IV$) characteristics and sample inductance with a resolution of a fraction of nH. The inductance is deduced from the center frequency shift of the RLC circuit resonance peak, see Figs.~\ref{fig:firstfig}\textbf{f},\textbf{g}.
The sample holder is mounted on the cold finger of a dilution refrigerator, and can be rotated \textit{in situ} via piezo-rotators. The rotation axis is such that the field of a superconducting coil always remains in the 2DEG plane. Two additional perpendicular coils allow us to accurately control the out-of-plane component $B_z$ of the magnetic field~(see Fig.~\ref{fig:firstfig}\textbf{a}).

\begin{figure*}[tb]
\includegraphics[width=\textwidth]{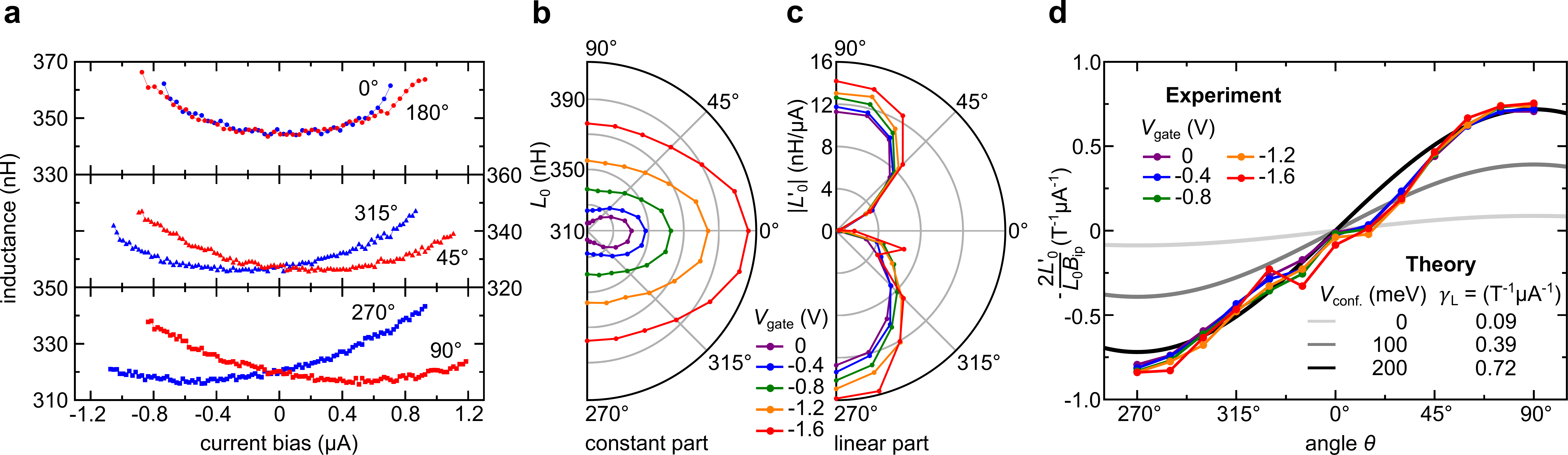}
\caption{\textbf{Supercurrent anisotropy and rectification}. \textbf{a},~Current dependence of the kinetic inductance of the array for an applied in-plane magnetic field of 100~mT. The top graphs refer to an in-plane field directed parallel to the current direction~$\hat{x}$. The middle graph refers to an in-plane magnetic field forming an angle of 45$^{\circ}$ with the current direction, while in the bottom graph the in-plane field is perpendicular to the current. Small vertical shifts have been applied to the curves to account for the residual field of the superconducting coil, as discussed in the Supplemental Material.  \textbf{b},~Constant and \textbf{c},~linear coefficients of the polynomial expansion of $L(I)$ as a function of the angle $\theta$ between $\vec{B}_\mathrm{ip}$ and the supercurrent density  oriented along $\hat{x}$. \textbf{d},~Color lines and symbols: measured supercurrent magnetochiral anisotropy $-2L'_0(T,\theta)/[L_0(T,\theta)B_\mathrm{ip}]$ vs.~magnetic-field angle $\theta$. 
A coefficient $\gamma_L=0.77\cdot 10^{6}\,$T$^{-1}$A$^{-1}$  is extracted from a sine fit of the data. 
Grey scale lines: fitted magnetochiral anisotropy  computed within our model for different values of the confinement potential $ V_\mathrm{conf}$. The corresponding values of the maximum magnetochiral anisotropy $ \gamma_L $ are reported.  The three curves are perfect sine functions. 
} 
\label{fig:asy}
\end{figure*}

As demonstrated in Ref.~\cite{baumgartner2020}, the 100~nm-long weak links in our samples are ballistic and close to the short-junction limit. The supercurrent is therefore carried by one Kramers pair of Andreev bound states~(ABS) per transverse mode. Their energy $\varepsilon_\mathrm{ABS}(\varphi)$ depends on the phase difference $\varphi$ between the superconducting leads. The phase~derivative of the ABS spectrum eventually defines the CPR, see black curve in Fig.~\ref{fig:firstfig}\textbf{d}. 
If \textit{either} time-reversal symmetry~(exchange of time direction) \textit{or} parity symmetry~(exchange of the leads) is preserved, then the ABS are symmetric, $\varepsilon_\mathrm{ABS}(\varphi)=\varepsilon_\mathrm{ABS}(-\varphi)$, and the CPR is antisymmetric, $I(\varphi)=-I(-\varphi)$, i.e., its Fourier expansion contains only sinusoidal terms. As an immediate consequence, the current is zero for zero phase difference, and vice versa. On the other hand, if \textit{both} symmetries are broken, the Kramers degeneracy between the two spin components is also broken, as shown in Fig.~\ref{fig:firstfig}\textbf{d} (red and blue curve). An anomalous CPR is thus observed~\cite{Bezuglyi2002,Buzdin2007,Reynoso2008,Reynoso2012,Yokoyama2014,Shen2014,Konschelle2015}, whose Fourier expansion contains also cosine~terms. In experiments, this is typically revealed by measuring the anomalous phase shift $\varphi_0$ of the CPR in phase-biased junctions~\cite{Szombati2016,Assouline2019,Mayer2020b}. 

If, however, the junction transparency is high \cite{Mayer2020b,baumgartner2020}, higher harmonic sine~terms appear in the CPR and the cosine~terms introduced by the in-plane magnetic field cannot be absorbed in a mere phase shift. In this case, the positive and negative current branches in the CPR can be markedly different, as illustrated in Fig.~\ref{fig:firstfig}\textbf{d}. This leads to two magnetoelectric effects: i)~the inflection point in the CPR occurs at a finite current $i^\ast$ that changes sign when reversing the in-plane magnetic field direction; ii)~the extremal values (i.e., the critical currents $ I^-_\mathrm{c} $ and $ I^+_\mathrm{c} $) for positive and negative phase difference differ, leading to a certain bias-current range in which a \textit{supercurrent diode} effect can be measured: 
for one current direction, superflow is observed ($ I<I^+_\mathrm{c} $), while for the other ($ |I|> |I^-_\mathrm{c}|$) the junctions are in their resistive state. This effect constitutes the Josephson junction analog of that recently reported by Ando~\textit{et al.}~\cite{Ando2020} in 2D metallic superlattices.
In what follows, we demonstrate both. 

\section{Magnetochiral anisotropy for supercurrents}

A convenient probe of the CPR symmetry is the Josephson inductance, which can be derived by combining the CPR $I=I_{\mathrm{c}0}f(\varphi)$ and the 2$^\mathrm{nd}$ Josephson equation $\dot{\varphi}=2eV/\hbar$ to
\begin{equation}
L\ =\ \frac{V}{{dI}/{dt}}\ =\ \frac {\hbar}{2eI_{\mathrm{c}0}f'(\varphi)}=\frac{\hbar}{2e}\frac{d\varphi(I)}{dI}.
\label{eq:JI}
\end{equation}
In other words, the Josephson inductance is  proportional to the derivative of the inverse CPR. Therefore, the minimum of $L(I)$ occurs at the inflection-point~current $i^\ast$ in the CPR~(red dot on blue curve in Fig.~\ref{fig:firstfig}\textbf{e}). As shown in the corresponding reference measurement in Fig.~\ref{fig:firstfig}\textbf{g}, $L(I)$ is symmetric around zero current, where the minimum inductance occurs in the absence of magnetic fields. The situation can change when an in-plane field is added, as shown in Fig.~\ref{fig:asy}\textbf{a}. If the in-plane field $\vec{B}_\mathrm{ip}=B_x\hat{x}+B_y\hat{y}$ is parallel to the current ($B_y=0$), no asymmetry is observed, see top panel in Fig.~\ref{fig:asy}\textbf{a}. The overall inductance increases, reflecting the gap (and thus the critical current) reduction, but no magnetochiral effect is observed, as the vector product $\vec{B}\times\vec{I}$ is still zero. On the contrary, when the in-plane field has a component $B_y$ perpendicular to the current, a clear asymmetry is observed: the minimum of $L(I)$~(corresponding to the inflection point in the CPR) occurs now at a finite current $i^\ast$. The value of $i^\ast$ increases with increasing $B_y$ and its sign switches together with the sign of $B_y$.

To quantify the effect, we take the leading terms in the polynomial expansion of $L(I)\approx L_0 + L_0'I+L_0''I^2/2$ around zero current, with $L_0'\equiv \partial_IL|_{I=0}$ and $L_0''\equiv \partial^2_IL|_{I=0}$. In Figs.~\ref{fig:asy}\textbf{b} and \ref{fig:asy}\textbf{c}, we plot the constant term $L_0$ and the linear term $L_0'$ as functions of the angle between the applied in-plane field $\vec{B}_\mathrm{ip}$ and the current direction $\hat{x}$. 
The constant term increases in magnitude when decreasing gate voltage towards more negative values and shows relatively small anisotropy. The increase of $L_0$ reflects the decrease of $I_{\mathrm{c}0}$ with decreasing number of channels. The slight anisotropy of $L_0$ probably reflects a warping of the Fermi surface in the parallel field that affects the Fermi~velocities of the two spin components. In contrast, the linear term is strongly anisotropic, as it vanishes for magnetic fields parallel to the current direction and reaches its maximum for transverse field~($B_x=0$). Very similar results have been found for a second sample with current flow in the [$1\overline{1}0$]~direction. Figure~\ref{fig:asy}\textbf{d} shows that the ratio $L^\prime_0/L_0$ is nearly independent of the gate voltage and varies in good approximation proportionally to $(\vec{B}\times\vec{I})\cdot\hat{z}=BI\sin \theta$, where $\theta$ is the angle between $\vec{B}_\mathrm{ip}$ and $\vec{I}$. From the amplitude of the sine, we extract the magnetochiral anisotropy coefficient for the inductance, $\gamma_L=0.76\cdot 10^6$~T$^{-1}$A$^{-1}$. This is a new observable that refers directly to the superfluid and cannot be detected by resistance measurements. Interestingly, it is of the same order of the corresponding coefficient for the resistance $\gamma_S$ discussed below, namely, in the range of $10^6$~T$^{-1}$A$^{-1}$.

\begin{figure*}[t]
	\includegraphics[width=\textwidth]{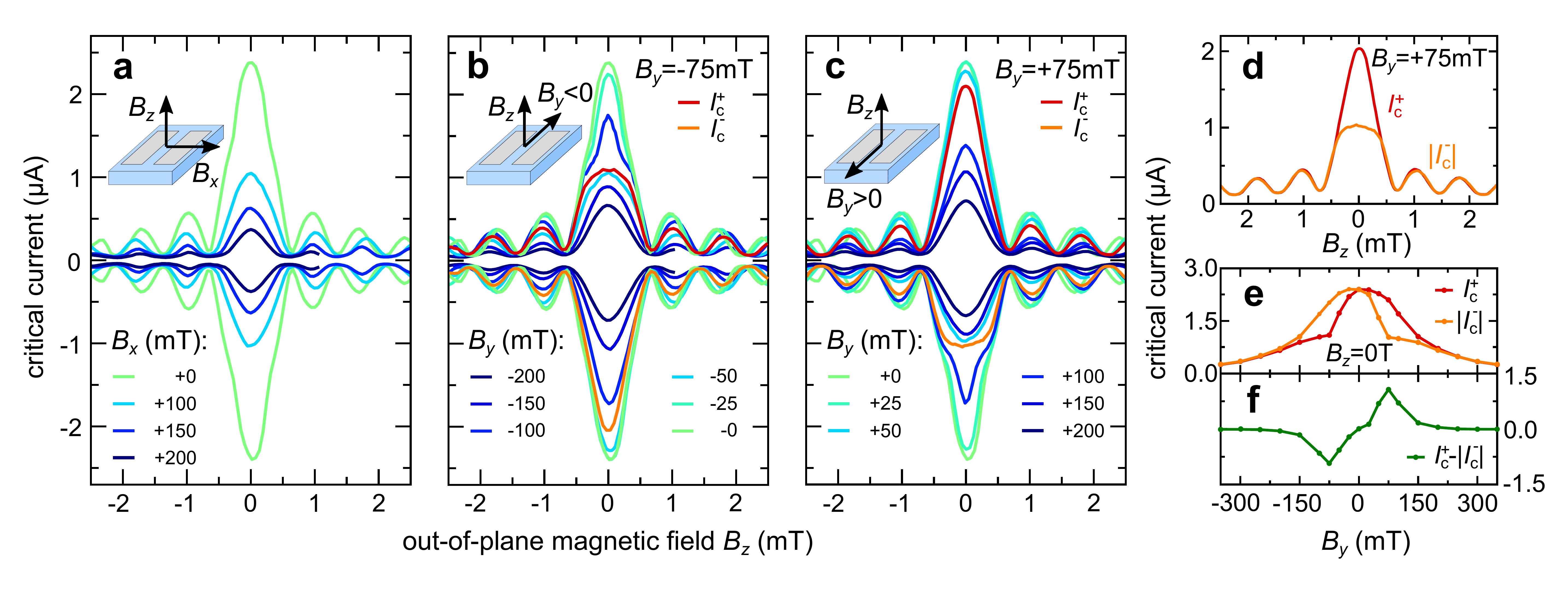}
	\caption{\textbf{Supercurrent interference}. \textbf{a},~Fraunhofer patterns of the Josephson junction array in combined out-of-plane $B_z$ and in-plane magnetic field $B_x$ parallel to the current flow, with $B_y=0$. Each curve represents the critical current $I_c(B_z)$ at a given value of $B_{x}$. \textbf{b},~Modified Fraunhofer patterns for in-plane fields $B_y<0$  transverse to the current flow, with $B_x=0$. \textbf{c},~The same as in panel \textbf{b}, but for $B_y>0$. \textbf{d},~Direct comparison of $I_\mathrm{c}^+$ and $|I_\mathrm{c}^-|$ vs. $B_{z}$ for $B_{y}=75$~mT, where the asymmetry is the largest. \textbf{e},~$I_\mathrm{c}^+$ and $|I_\mathrm{c}^-|$ as a function of $B_{y}$  for $B_{z}=0$. \textbf{f},~Difference between $I_\mathrm{c}^+$ and $|I_\mathrm{c}^-|$ for the latter case.}
	\label{fig:fraun}
\end{figure*}

        To numerically simulate the $ L(I) $-characteristics of the  Josephson junction~array and to extract the supercurrent magnetochiral anisotropy coefficient, we developed a realistic theoretical model~\footnote{See Supplemental Material for further information.} whose \textsc{Kwant}~\cite{Groth2014} implementation provides a semi-quantitative description of our experimental data. The fundamental parameters in our model are the Bychkov--Rashba spin-orbit coupling strength~$ \alpha = 15 \, \mathrm{meV} \, \mathrm{nm} $~\cite{Note1,Mayer2020}, the in-plane $ g $-factor $ g^* \approx -10 $ of the InAs quantum well~\cite{Mayer2020}, the effective~mass of electrons $ m^* \approx 0.02 m_0 $~($ m_0 $ is the free-electron~mass)~\cite{Vurgaftman2001,Fabian2007}, as well as the Fermi~energy that we estimated as~$ \mu \approx 239 \, \mathrm{meV} $. 
        We approximate the effects of charge transfer and band alignment   within the InAs~layer by a parabolic potential~well of depth~$ V_\mathrm{conf} $. 
        This parameter controls the magnitude of the anisotropy with respect to the orientation of the in-plane magnetic field. 
        The ratio of the Zeeman shift of the two spin subbands   to the Fermi energy $\mu$ controls the orbital phase shift between the subbands forming the ABS~\cite{Reynoso2008,Reynoso2012,Yokoyama2014}. Hence, the Zeeman shifts become more important at low $\mu$ (and thus high $ V_\mathrm{conf} $). The lower the Fermi level, the lower the $ \hat{x} $-component of the Fermi velocity, and thus the higher the CPR asymmetry~\cite{Buzdin2008}. As discussed in the Supplemental Material, the effect is amplified in the presence of many transverse channels in wide junctions: modes with large transverse wavevector $k_{\mathrm{F},y}$ must have a low longitudinal $k_{\mathrm{F},x}$. These modes contribute overproportionally to the anomalous CPR shift $\varphi_0$~\cite{Mayer2020b}.    
        
 The results of our numerical simulations are also shown in~Fig.~\ref{fig:asy}\textbf{d} and fully support our qualitative reasoning. 
        In particular, the sinusoidal dependence of $L^\prime_0(\theta)$ on the in-plane angle $\theta$ is nicely reproduced.
        Assuming $V_\mathrm{conf} = 0 $, we obtain supercurrent magnetochiral anisotropy coefficients $\gamma_L$ that are about one order of magnitude lower than the experimentally detected ones.
 Since the anisotropy increases linearly with the Rashba~parameter~(analogously to~$ \varphi_0 $~\cite{Buzdin2008,Note1}), reproducing the experimental values would require its enhancement   by approximately one order of magnitude, which is unrealistic. 
        Instead, we assume a finite confinement potential $V_\mathrm{conf}$, which we estimate to be about $200 \, \mathrm{meV} $ from the comparison with the measured data. 
        This value is not far from the $ 150 \, \mathrm{meV} $ value reported in an earlier work for the confinement in symmetric AlGa/GaAs/AlGa~multilayers~\cite{Seraide2002}. Importantly, our model reproduces within a factor 4 the normal state Sharvin resistance of the channel.  This consistency indicates that the measured value of $\gamma_L$, the novel quantity introduced in this work, can be justified at a microscopic level using reasonable material parameters.


Concluding this section, the kinetic inductance reveals a magnetochiral anisotropy \textit{deep in the superconducting state}, which is consistent with Eq.~\ref{eq:gammaell}.  
The asymmetry with respect to the current is ultimately produced by the combination of orthogonal electric~(leading to Rashba SOI) and magnetic fields that are both perpendicular to the current direction. Therefore, this is the \textit{superfluid analog} of the magnetochiral effect for the resistance observed in the fluctuation regime of noncentrosymmetric superconductors~\cite{Wakatsuki2017,Itahashi2020}.  
We emphasize that, despite the similarities between Eqs.~\ref{eq:rikken} and~\ref{eq:gammaell}, the important difference between the two magnetoelectric effects is that the former relies on dissipation, while the latter can instead be measured everywhere in the superconducting phase.


\section{Josephson junction rectifiers}
In order to check for the second magnetoelectric effect---the anomalous critical currents $I_\mathrm{c}^+$ and $I_\mathrm{c}^-$---we have measured supercurrent-interference patterns by applying a small out-of-plane magnetic field $\vec{B}_z\parallel \hat{z}$ coexisting with the in-plane field $\vec{B}_\mathrm{ip}= B_x \hat{x}+B_y \hat{y}$, whose magnitude and orientation with respect to the current direction can be controlled. The results are shown in Fig.~\ref{fig:fraun}. 
The different curves show the positive $I_\mathrm{c}^+$ and negative $I_\mathrm{c}^-$  critical currents displayed as a function of the out-of-plane magnetic field  in the upper and lower half-planes, respectively. 
For both current orientation, the DC bias  was swept from zero to finite  (positive or negative) values. The critical current was determined as the current bias producing a threshold voltage of roughly 1~\textmu V per junction (2~mV in total). In this way, a heating-induced hysteresis of the switching current is excluded.

In the absence of in-plane fields, the pattern is symmetric and Fraunhofer-like~\cite{Suominen2017,Mayer2020b,Guiducci2019} in the current direction (Fig.~\ref{fig:fraun}\textbf{a}). The introduction of an in-plane field parallel to the current reduces the gap and thus the critical current, but does not introduce any asymmetry between the positive and negative current directions. Besides some variations of the apparent period of the patterns with $B_\mathrm{ip}$, the lobe structure approaches the standard Fraunhofer shape.


On the other hand, in the presence of a transverse in-plane field component $B_y$ (perpendicular to the current), a clear asymmetry between positive and negative current bias is produced. Data in Figs.~\ref{fig:fraun}\textbf{b}, \textbf{c} suggest that the effect is noticeable only for the main lobe of the diffraction pattern.  A striking consequence of such asymmetry is the  superconducting diode effect, which manifests itself in the difference between the critical currents ($|I_\mathrm{c}^+|$ and $|I_\mathrm{c}^-|$) corresponding to the two current polarities. In the current range between $|I_\mathrm{c}^-|$ and $|I_\mathrm{c}^+|$, supercurrent can only flow in one direction, controllable through the in-plane magnetic field~\cite{Ando2020}. To highlight the diode effect, we select the diffraction patterns for $B_y=+75$~mT and both current polarities $I_\mathrm{c}^+$ and $I_\mathrm{c}^-$ from panel~\textbf{c}, see the red and orange curves in Fig.~\ref{fig:fraun}\textbf{d}. Figure~\ref{fig:fraun}\textbf{e} shows the corresponding data vs.~transverse in-plane field $B_y$ (red and orange curves).
Interestingly, the diode effect is pronounced only for a finite range of in-plane field magnitudes, where the difference $I_\mathrm{c}^+-|I_\mathrm{c}^-|$  (green curve in Fig.~\ref{fig:fraun}\textbf{f}) is sizable.
This can be understood as follows: if the transverse field $B_y$ is too low, the resulting symmetry breaking is too weak to produce a significant asymmetry. On the contrary, if $B_y$ is too large, the ratio $\Delta^\ast/k_\mathrm{B}T$ of induced gap and temperature is reduced, and the CPR approaches a sinusoidal shape. In this case, the SOI~effect is reduced to a pure $\varphi_0$-phase shift of the CPR and the difference $I_\mathrm{c}^+-|I_\mathrm{c}^-|$ is suppressed.
A detailed modeling of the combined $B_z$- and $B_y$-dependence must include not only supercurrent interference, but also the orbital pair-breaking in the proximitized heterostructure, and is beyond the scope of this work.

\section{Magnetochiral anisotropy of the resistance}

\begin{figure*}[t]
\includegraphics[width=\textwidth]{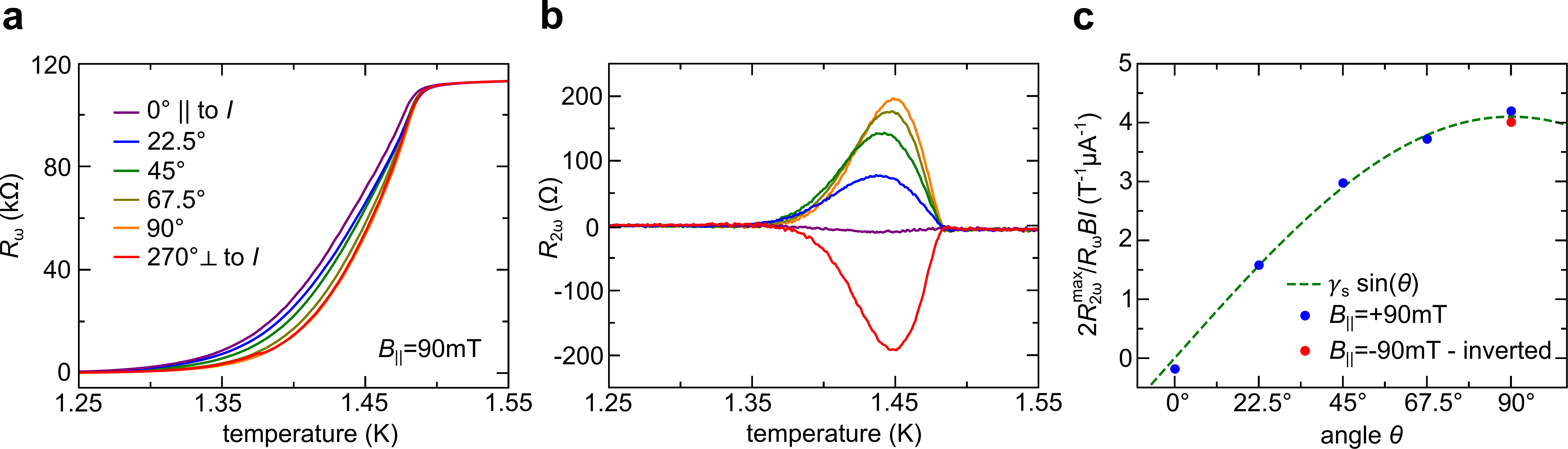}
\caption{\textbf{Magnetochiral anisotropy in the fluctuation regime}. \textbf{a},~Resistive transition $R_\omega(T,\theta)$ for different angles $\theta$ of the in-plane magnetic field. This measurement has been performed on another array. \textbf{b},~Second harmonics $R_{2\omega}(T,\theta)=V_{2\omega}(T,\theta)/I_\mathrm{ac}$ of the $V(I)$-characteristics. \textbf{c},~Fluctuation magnetochiral anisotropy $2R_{2\omega}(T,\theta)/R_{\omega}(T,\theta)$ vs.~in-plane angle $\theta$. The coefficient $\gamma_S=4.1\cdot 10^{6}\,$T$^{-1}$A$^{-1}$ is extracted from a sine fit of the data.} 
\label{fig:mca}
\end{figure*}

It is interesting to check whether our Josephson junctions also display a magnetochiral anisotropy of the resistance in the regime of thermal phase fluctuations close to $T_\mathrm{c}$. A convenient way to measure the effect~\cite{Wakatsuki2017,Itahashi2020,Ideue2017,He2018} is by lock-in techniques. A linear variation of the differential resistance on the current~[and thus a quadratic term in the voltage--current characteristic $V(I)$] can be detected measuring the $2\omega$ voltage response to a sinusoidal current excitation at frequency $\omega$. We expect the effect to be negligible both at temperatures well above $T_\mathrm{c}$~($\gamma_N\ll\gamma_S$) and below $T_\mathrm{c}$~(no measurable resistance, $R_0=0$).  Figures~\ref{fig:mca}\textbf{a}~and~\textbf{b} show, respectively, the linear~($R_{\omega}$) and the quadratic~($R_{2\omega}$) contribution to $V(I)$ as a function of temperature and for different in-plane fields. As expected, we observe a finite magnetochiral anisotropy coefficient $\gamma_S\simeq 4.1\cdot10^6$~T$^{-1}$A$^{-1}$ near the transition temperature. 
The product between $\gamma_S$ and sample width is about 12.9~T$^{-1}$A$^{-1}$m, similar~\cite{Note1} to that recently reported in the {phase} fluctuation regime of 2D interfacial superconductors~\cite{Itahashi2020}.  As shown in Fig.~\ref{fig:mca}\textbf{c}, the nonlinear resistance is proportional to the field component perpendicular to the current direction. By rotating the sample with respect to the magnetic field, we clearly see a sinusoidal variation with the angle $\theta$ between current and field, as expected from Eq.~\ref{eq:rikken}.


In conclusion, we have demonstrated that Josephson junctions with strong spin-orbit interactions display a strong supercurrent magnetochiral anisotropy and a corresponding supercurrent diode effect. We have characterized the anisotropy coefficient both deep in the superconducting regime and in the phase-fluctuation regime, and have found it to be comparable to the values reported in the phase-fluctuation regime of 2D superconductors. Spatially symmetric Josephson junctions act as controllable supercurrent rectifiers, which may find applications in microwave quantum electronic circuits.


\section*{Acknowledgments}
\begin{acknowledgments}
    Work at Regensburg University was funded by the Deutsche Forschungsgemeinschaft (DFG, German Research Foundation) – Project-ID 314695032 – SFB 1277 (Subprojects B05, B07, and B08). The theory part also benefited from the European Union’s Horizon 2020 research and innovation programme under Grant Agreement No.~881603~(Graphene Flagship Core~3). Work completed at Purdue University is supported by Microsoft Quantum. 
    A.C. thanks Michael~Barth for valuable discussions on \textsc{Kwant}'s functionalities.
\end{acknowledgments}
\vspace{2mm}

 \noindent {\it Author Contributions}: 
            $^\dagger$C.~Baumgartner and $^\dagger$L.~Fuchs contributed equally to this work. C.B. fabricated the device, and performed the measurements.  L.F., and S.R. developed and optimized the measurement method.   T.L., S.G., and, G.C.G.~designed the heterostructure, conducted MBE growth, and performed initial characterization of the hybrid superconductor/semiconductor wafer. C.B. and N.P. analyzed the data. N.P. and C.S. conceived the experiment. A.C., D.K., and J.F. formulated the theoretical model.   A.C. performed \textsc{Kwant} simulations, P.E.F.Jr. conducted the $ \mathbf{k} \cdot \mathbf{p}$-calculations,  C.S., J.F., and M.J.M supervised research activities at Regensburg and Purdue, respectively. All authors contributed to discussions and the writing of the manuscript.

\bibliography{biblio}

\begin{thebibliography}{60}%
\makeatletter
\providecommand \@ifxundefined [1]{%
 \@ifx{#1\undefined}
}%
\providecommand \@ifnum [1]{%
 \ifnum #1\expandafter \@firstoftwo
 \else \expandafter \@secondoftwo
 \fi
}%
\providecommand \@ifx [1]{%
 \ifx #1\expandafter \@firstoftwo
 \else \expandafter \@secondoftwo
 \fi
}%
\providecommand \natexlab [1]{#1}%
\providecommand \enquote  [1]{``#1''}%
\providecommand \bibnamefont  [1]{#1}%
\providecommand \bibfnamefont [1]{#1}%
\providecommand \citenamefont [1]{#1}%
\providecommand \href@noop [0]{\@secondoftwo}%
\providecommand \href [0]{\begingroup \@sanitize@url \@href}%
\providecommand \@href[1]{\@@startlink{#1}\@@href}%
\providecommand \@@href[1]{\endgroup#1\@@endlink}%
\providecommand \@sanitize@url [0]{\catcode `\\12\catcode `\$12\catcode
  `\&12\catcode `\#12\catcode `\^12\catcode `\_12\catcode `\%12\relax}%
\providecommand \@@startlink[1]{}%
\providecommand \@@endlink[0]{}%
\providecommand \url  [0]{\begingroup\@sanitize@url \@url }%
\providecommand \@url [1]{\endgroup\@href {#1}{\urlprefix }}%
\providecommand \urlprefix  [0]{URL }%
\providecommand \Eprint [0]{\href }%
\providecommand \doibase [0]{https://doi.org/}%
\providecommand \selectlanguage [0]{\@gobble}%
\providecommand \bibinfo  [0]{\@secondoftwo}%
\providecommand \bibfield  [0]{\@secondoftwo}%
\providecommand \translation [1]{[#1]}%
\providecommand \BibitemOpen [0]{}%
\providecommand \bibitemStop [0]{}%
\providecommand \bibitemNoStop [0]{.\EOS\space}%
\providecommand \EOS [0]{\spacefactor3000\relax}%
\providecommand \BibitemShut  [1]{\csname bibitem#1\endcsname}%
\let\auto@bib@innerbib\@empty
\bibitem [{\citenamefont {Scaff}\ and\ \citenamefont
  {Ohl}(1947)}]{ScaffOhl1947}%
  \BibitemOpen
  \bibfield  {author} {\bibinfo {author} {\bibfnamefont {J.~H.}\ \bibnamefont
  {Scaff}}\ and\ \bibinfo {author} {\bibfnamefont {R.~S.}\ \bibnamefont
  {Ohl}},\ }\bibfield  {title} {\bibinfo {title} {{Development of Silicon
  Crystal Rectifiers for Microwave Radar Receivers}},\ }\href
  {https://doi.org/https://doi.org/10.1002/j.1538-7305.1947.tb01310.x}
  {\bibfield  {journal} {\bibinfo  {journal} {Bell System Technical Journal}\
  }\textbf {\bibinfo {volume} {26}},\ \bibinfo {pages} {1} (\bibinfo {year}
  {1947})}\BibitemShut {NoStop}%
\bibitem [{\citenamefont {{Shockley}}(1949)}]{Shockley1949}%
  \BibitemOpen
  \bibfield  {author} {\bibinfo {author} {\bibfnamefont {W.}~\bibnamefont
  {{Shockley}}},\ }\bibfield  {title} {\bibinfo {title} {{The theory of p-n
  junctions in semiconductors and p-n junction transistors}},\ }\href
  {https://doi.org/10.1002/j.1538-7305.1949.tb03645.x} {\bibfield  {journal}
  {\bibinfo  {journal} {The Bell System Technical Journal}\ }\textbf {\bibinfo
  {volume} {28}},\ \bibinfo {pages} {435} (\bibinfo {year} {1949})}\BibitemShut
  {NoStop}%
\bibitem [{\citenamefont {Onsager}(1931)}]{OnsagerPR31}%
  \BibitemOpen
  \bibfield  {author} {\bibinfo {author} {\bibfnamefont {L.}~\bibnamefont
  {Onsager}},\ }\bibfield  {title} {\bibinfo {title} {{Reciprocal Relations in
  Irreversible Processes. I.}},\ }\href
  {https://doi.org/10.1103/PhysRev.37.405} {\bibfield  {journal} {\bibinfo
  {journal} {Phys. Rev.}\ }\textbf {\bibinfo {volume} {37}},\ \bibinfo {pages}
  {405} (\bibinfo {year} {1931})}\BibitemShut {NoStop}%
\bibitem [{\citenamefont {Kubo}(1957)}]{Kubo57}%
  \BibitemOpen
  \bibfield  {author} {\bibinfo {author} {\bibfnamefont {R.}~\bibnamefont
  {Kubo}},\ }\bibfield  {title} {\bibinfo {title} {{Statistical-Mechanical
  Theory of Irreversible Processes. I. General Theory and Simple Applications
  to Magnetic and Conduction Problems}},\ }\href
  {https://doi.org/10.1143/JPSJ.12.570} {\bibfield  {journal} {\bibinfo
  {journal} {Journal of the Physical Society of Japan}\ }\textbf {\bibinfo
  {volume} {12}},\ \bibinfo {pages} {570} (\bibinfo {year} {1957})}\BibitemShut
  {NoStop}%
\bibitem [{\citenamefont {Tokura}\ and\ \citenamefont
  {Nagaosa}(2018)}]{Tokura2018}%
  \BibitemOpen
  \bibfield  {author} {\bibinfo {author} {\bibfnamefont {Y.}~\bibnamefont
  {Tokura}}\ and\ \bibinfo {author} {\bibfnamefont {N.}~\bibnamefont
  {Nagaosa}},\ }\bibfield  {title} {\bibinfo {title} {{Nonreciprocal responses
  from non-centrosymmetric quantum materials}},\ }\href
  {https://doi.org/10.1038/s41467-018-05759-4} {\bibfield  {journal} {\bibinfo
  {journal} {Nature Communications}\ }\textbf {\bibinfo {volume} {9}},\
  \bibinfo {pages} {3740} (\bibinfo {year} {2018})}\BibitemShut {NoStop}%
\bibitem [{\citenamefont {Hoshino}\ \emph {et~al.}(2018)\citenamefont
  {Hoshino}, \citenamefont {Wakatsuki}, \citenamefont {Hamamoto},\ and\
  \citenamefont {Nagaosa}}]{Hoshino2018}%
  \BibitemOpen
  \bibfield  {author} {\bibinfo {author} {\bibfnamefont {S.}~\bibnamefont
  {Hoshino}}, \bibinfo {author} {\bibfnamefont {R.}~\bibnamefont {Wakatsuki}},
  \bibinfo {author} {\bibfnamefont {K.}~\bibnamefont {Hamamoto}},\ and\
  \bibinfo {author} {\bibfnamefont {N.}~\bibnamefont {Nagaosa}},\ }\bibfield
  {title} {\bibinfo {title} {{Nonreciprocal charge transport in two-dimensional
  noncentrosymmetric superconductors}},\ }\href
  {https://doi.org/10.1103/PhysRevB.98.054510} {\bibfield  {journal} {\bibinfo
  {journal} {Phys. Rev. B}\ }\textbf {\bibinfo {volume} {98}},\ \bibinfo
  {pages} {054510} (\bibinfo {year} {2018})}\BibitemShut {NoStop}%
\bibitem [{\citenamefont {Rikken}\ \emph {et~al.}(2001)\citenamefont {Rikken},
  \citenamefont {F\"olling},\ and\ \citenamefont {Wyder}}]{RikkenPRL2001}%
  \BibitemOpen
  \bibfield  {author} {\bibinfo {author} {\bibfnamefont {G.~L. J.~A.}\
  \bibnamefont {Rikken}}, \bibinfo {author} {\bibfnamefont {J.}~\bibnamefont
  {F\"olling}},\ and\ \bibinfo {author} {\bibfnamefont {P.}~\bibnamefont
  {Wyder}},\ }\bibfield  {title} {\bibinfo {title} {{Electrical Magnetochiral
  Anisotropy}},\ }\href {https://doi.org/10.1103/PhysRevLett.87.236602}
  {\bibfield  {journal} {\bibinfo  {journal} {Phys. Rev. Lett.}\ }\textbf
  {\bibinfo {volume} {87}},\ \bibinfo {pages} {236602} (\bibinfo {year}
  {2001})}\BibitemShut {NoStop}%
\bibitem [{\citenamefont {Rikken}\ and\ \citenamefont
  {Wyder}(2005)}]{RikkenPRL2005}%
  \BibitemOpen
  \bibfield  {author} {\bibinfo {author} {\bibfnamefont {G.~L. J.~A.}\
  \bibnamefont {Rikken}}\ and\ \bibinfo {author} {\bibfnamefont
  {P.}~\bibnamefont {Wyder}},\ }\bibfield  {title} {\bibinfo {title}
  {{Magnetoelectric Anisotropy in Diffusive Transport}},\ }\href
  {https://doi.org/10.1103/PhysRevLett.94.016601} {\bibfield  {journal}
  {\bibinfo  {journal} {Phys. Rev. Lett.}\ }\textbf {\bibinfo {volume} {94}},\
  \bibinfo {pages} {016601} (\bibinfo {year} {2005})}\BibitemShut {NoStop}%
\bibitem [{\citenamefont {Wakatsuki}\ \emph {et~al.}(2017)\citenamefont
  {Wakatsuki}, \citenamefont {Saito}, \citenamefont {Hoshino}, \citenamefont
  {Itahashi}, \citenamefont {Ideue}, \citenamefont {Ezawa}, \citenamefont
  {Iwasa},\ and\ \citenamefont {Nagaosa}}]{Wakatsuki2017}%
  \BibitemOpen
  \bibfield  {author} {\bibinfo {author} {\bibfnamefont {R.}~\bibnamefont
  {Wakatsuki}}, \bibinfo {author} {\bibfnamefont {Y.}~\bibnamefont {Saito}},
  \bibinfo {author} {\bibfnamefont {S.}~\bibnamefont {Hoshino}}, \bibinfo
  {author} {\bibfnamefont {Y.~M.}\ \bibnamefont {Itahashi}}, \bibinfo {author}
  {\bibfnamefont {T.}~\bibnamefont {Ideue}}, \bibinfo {author} {\bibfnamefont
  {M.}~\bibnamefont {Ezawa}}, \bibinfo {author} {\bibfnamefont
  {Y.}~\bibnamefont {Iwasa}},\ and\ \bibinfo {author} {\bibfnamefont
  {N.}~\bibnamefont {Nagaosa}},\ }\bibfield  {title} {\bibinfo {title}
  {{Nonreciprocal charge transport in noncentrosymmetric superconductors}},\
  }\href@noop {} {\bibfield  {journal} {\bibinfo  {journal} {Science Advances}\
  }\textbf {\bibinfo {volume} {3}},\ \bibinfo {pages} {e1602390} (\bibinfo
  {year} {2017})}\BibitemShut {NoStop}%
\bibitem [{\citenamefont {Itahashi}\ \emph {et~al.}(2020)\citenamefont
  {Itahashi}, \citenamefont {Ideue}, \citenamefont {Saito}, \citenamefont
  {Shimizu}, \citenamefont {Ouchi}, \citenamefont {Nojima},\ and\ \citenamefont
  {Iwasa}}]{Itahashi2020}%
  \BibitemOpen
  \bibfield  {author} {\bibinfo {author} {\bibfnamefont {Y.~M.}\ \bibnamefont
  {Itahashi}}, \bibinfo {author} {\bibfnamefont {T.}~\bibnamefont {Ideue}},
  \bibinfo {author} {\bibfnamefont {Y.}~\bibnamefont {Saito}}, \bibinfo
  {author} {\bibfnamefont {S.}~\bibnamefont {Shimizu}}, \bibinfo {author}
  {\bibfnamefont {T.}~\bibnamefont {Ouchi}}, \bibinfo {author} {\bibfnamefont
  {T.}~\bibnamefont {Nojima}},\ and\ \bibinfo {author} {\bibfnamefont
  {Y.}~\bibnamefont {Iwasa}},\ }\bibfield  {title} {\bibinfo {title}
  {{Nonreciprocal transport in gate-induced polar superconductor SrTiO$_3$}},\
  }\href@noop {} {\bibfield  {journal} {\bibinfo  {journal} {Science Advances}\
  }\textbf {\bibinfo {volume} {6}},\ \bibinfo {pages} {eaay9120} (\bibinfo
  {year} {2020})}\BibitemShut {NoStop}%
\bibitem [{\citenamefont {Baumgartner}\ \emph {et~al.}(2021)\citenamefont
  {Baumgartner}, \citenamefont {Fuchs}, \citenamefont {Fr\'esz}, \citenamefont
  {Reinhardt}, \citenamefont {Gronin}, \citenamefont {Gardner}, \citenamefont
  {Manfra}, \citenamefont {Paradiso},\ and\ \citenamefont
  {Strunk}}]{baumgartner2020}%
  \BibitemOpen
  \bibfield  {author} {\bibinfo {author} {\bibfnamefont {C.}~\bibnamefont
  {Baumgartner}}, \bibinfo {author} {\bibfnamefont {L.}~\bibnamefont {Fuchs}},
  \bibinfo {author} {\bibfnamefont {L.}~\bibnamefont {Fr\'esz}}, \bibinfo
  {author} {\bibfnamefont {S.}~\bibnamefont {Reinhardt}}, \bibinfo {author}
  {\bibfnamefont {S.}~\bibnamefont {Gronin}}, \bibinfo {author} {\bibfnamefont
  {G.~C.}\ \bibnamefont {Gardner}}, \bibinfo {author} {\bibfnamefont {M.~J.}\
  \bibnamefont {Manfra}}, \bibinfo {author} {\bibfnamefont {N.}~\bibnamefont
  {Paradiso}},\ and\ \bibinfo {author} {\bibfnamefont {C.}~\bibnamefont
  {Strunk}},\ }\bibfield  {title} {\bibinfo {title} {{Josephson Inductance as a
  Probe for Highly Ballistic Semiconductor-Superconductor Weak Links}},\ }\href
  {https://doi.org/10.1103/PhysRevLett.126.037001} {\bibfield  {journal}
  {\bibinfo  {journal} {Phys. Rev. Lett.}\ }\textbf {\bibinfo {volume} {126}},\
  \bibinfo {pages} {037001} (\bibinfo {year} {2021})}\BibitemShut {NoStop}%
\bibitem [{\citenamefont {Bezuglyi}\ \emph {et~al.}(2002)\citenamefont
  {Bezuglyi}, \citenamefont {Rozhavsky}, \citenamefont {Vagner},\ and\
  \citenamefont {Wyder}}]{Bezuglyi2002}%
  \BibitemOpen
  \bibfield  {author} {\bibinfo {author} {\bibfnamefont {E.~V.}\ \bibnamefont
  {Bezuglyi}}, \bibinfo {author} {\bibfnamefont {A.~S.}\ \bibnamefont
  {Rozhavsky}}, \bibinfo {author} {\bibfnamefont {I.~D.}\ \bibnamefont
  {Vagner}},\ and\ \bibinfo {author} {\bibfnamefont {P.}~\bibnamefont
  {Wyder}},\ }\bibfield  {title} {\bibinfo {title} {{Combined effect of Zeeman
  splitting and spin-orbit interaction on the Josephson current in a
  superconductor--two-dimensional electron gas--superconductor structure}},\
  }\href {https://doi.org/10.1103/PhysRevB.66.052508} {\bibfield  {journal}
  {\bibinfo  {journal} {Phys. Rev. B}\ }\textbf {\bibinfo {volume} {66}},\
  \bibinfo {pages} {052508} (\bibinfo {year} {2002})}\BibitemShut {NoStop}%
\bibitem [{\citenamefont {Buzdin}(2008{\natexlab{a}})}]{Buzdin2007}%
  \BibitemOpen
  \bibfield  {author} {\bibinfo {author} {\bibfnamefont {A.}~\bibnamefont
  {Buzdin}},\ }\bibfield  {title} {\bibinfo {title} {{Direct Coupling Between
  Magnetism and Superconducting Current in the Josephson $\varphi_0$
  junction}},\ }\href {https://doi.org/10.1103/PhysRevLett.101.107005}
  {\bibfield  {journal} {\bibinfo  {journal} {Phys. Rev. Lett.}\ }\textbf
  {\bibinfo {volume} {101}},\ \bibinfo {pages} {107005} (\bibinfo {year}
  {2008}{\natexlab{a}})}\BibitemShut {NoStop}%
\bibitem [{\citenamefont {Reynoso}\ \emph {et~al.}(2008)\citenamefont
  {Reynoso}, \citenamefont {Usaj}, \citenamefont {Balseiro}, \citenamefont
  {Feinberg},\ and\ \citenamefont {Avignon}}]{Reynoso2008}%
  \BibitemOpen
  \bibfield  {author} {\bibinfo {author} {\bibfnamefont {A.~A.}\ \bibnamefont
  {Reynoso}}, \bibinfo {author} {\bibfnamefont {G.}~\bibnamefont {Usaj}},
  \bibinfo {author} {\bibfnamefont {C.~A.}\ \bibnamefont {Balseiro}}, \bibinfo
  {author} {\bibfnamefont {D.}~\bibnamefont {Feinberg}},\ and\ \bibinfo
  {author} {\bibfnamefont {M.}~\bibnamefont {Avignon}},\ }\bibfield  {title}
  {\bibinfo {title} {{Anomalous Josephson Current in Junctions with Spin
  Polarizing Quantum Point Contacts}},\ }\href
  {https://doi.org/10.1103/PhysRevLett.101.107001} {\bibfield  {journal}
  {\bibinfo  {journal} {Phys. Rev. Lett.}\ }\textbf {\bibinfo {volume} {101}},\
  \bibinfo {pages} {107001} (\bibinfo {year} {2008})}\BibitemShut {NoStop}%
\bibitem [{\citenamefont {Reynoso}\ \emph {et~al.}(2012)\citenamefont
  {Reynoso}, \citenamefont {Usaj}, \citenamefont {Balseiro}, \citenamefont
  {Feinberg},\ and\ \citenamefont {Avignon}}]{Reynoso2012}%
  \BibitemOpen
  \bibfield  {author} {\bibinfo {author} {\bibfnamefont {A.~A.}\ \bibnamefont
  {Reynoso}}, \bibinfo {author} {\bibfnamefont {G.}~\bibnamefont {Usaj}},
  \bibinfo {author} {\bibfnamefont {C.~A.}\ \bibnamefont {Balseiro}}, \bibinfo
  {author} {\bibfnamefont {D.}~\bibnamefont {Feinberg}},\ and\ \bibinfo
  {author} {\bibfnamefont {M.}~\bibnamefont {Avignon}},\ }\bibfield  {title}
  {\bibinfo {title} {{Spin-orbit-induced chirality of Andreev states in
  Josephson junctions}},\ }\href {https://doi.org/10.1103/PhysRevB.86.214519}
  {\bibfield  {journal} {\bibinfo  {journal} {Phys. Rev. B}\ }\textbf {\bibinfo
  {volume} {86}},\ \bibinfo {pages} {214519} (\bibinfo {year}
  {2012})}\BibitemShut {NoStop}%
\bibitem [{\citenamefont {Yokoyama}\ \emph {et~al.}(2014)\citenamefont
  {Yokoyama}, \citenamefont {Eto},\ and\ \citenamefont
  {Nazarov}}]{Yokoyama2014}%
  \BibitemOpen
  \bibfield  {author} {\bibinfo {author} {\bibfnamefont {T.}~\bibnamefont
  {Yokoyama}}, \bibinfo {author} {\bibfnamefont {M.}~\bibnamefont {Eto}},\ and\
  \bibinfo {author} {\bibfnamefont {Y.~V.}\ \bibnamefont {Nazarov}},\
  }\bibfield  {title} {\bibinfo {title} {{Anomalous Josephson effect induced by
  spin-orbit interaction and Zeeman effect in semiconductor nanowires}},\
  }\href {https://doi.org/10.1103/PhysRevB.89.195407} {\bibfield  {journal}
  {\bibinfo  {journal} {Phys. Rev. B}\ }\textbf {\bibinfo {volume} {89}},\
  \bibinfo {pages} {195407} (\bibinfo {year} {2014})}\BibitemShut {NoStop}%
\bibitem [{\citenamefont {Shen}\ \emph {et~al.}(2014)\citenamefont {Shen},
  \citenamefont {Vignale},\ and\ \citenamefont {Raimondi}}]{Shen2014}%
  \BibitemOpen
  \bibfield  {author} {\bibinfo {author} {\bibfnamefont {K.}~\bibnamefont
  {Shen}}, \bibinfo {author} {\bibfnamefont {G.}~\bibnamefont {Vignale}},\ and\
  \bibinfo {author} {\bibfnamefont {R.}~\bibnamefont {Raimondi}},\ }\bibfield
  {title} {\bibinfo {title} {{Microscopic Theory of the Inverse Edelstein
  Effect}},\ }\href {https://doi.org/10.1103/PhysRevLett.112.096601} {\bibfield
   {journal} {\bibinfo  {journal} {Phys. Rev. Lett.}\ }\textbf {\bibinfo
  {volume} {112}},\ \bibinfo {pages} {096601} (\bibinfo {year}
  {2014})}\BibitemShut {NoStop}%
\bibitem [{\citenamefont {Konschelle}\ \emph {et~al.}(2015)\citenamefont
  {Konschelle}, \citenamefont {Tokatly},\ and\ \citenamefont
  {Bergeret}}]{Konschelle2015}%
  \BibitemOpen
  \bibfield  {author} {\bibinfo {author} {\bibfnamefont {F.}~\bibnamefont
  {Konschelle}}, \bibinfo {author} {\bibfnamefont {I.~V.}\ \bibnamefont
  {Tokatly}},\ and\ \bibinfo {author} {\bibfnamefont {F.~S.}\ \bibnamefont
  {Bergeret}},\ }\bibfield  {title} {\bibinfo {title} {{Theory of the
  spin-galvanic effect and the anomalous phase shift
  ${\ensuremath{\varphi}}_{0}$ in superconductors and Josephson junctions with
  intrinsic spin-orbit coupling}},\ }\href
  {https://doi.org/10.1103/PhysRevB.92.125443} {\bibfield  {journal} {\bibinfo
  {journal} {Phys. Rev. B}\ }\textbf {\bibinfo {volume} {92}},\ \bibinfo
  {pages} {125443} (\bibinfo {year} {2015})}\BibitemShut {NoStop}%
\bibitem [{\citenamefont {Szombati}\ \emph {et~al.}(2016)\citenamefont
  {Szombati}, \citenamefont {Nadj-Perge}, \citenamefont {Car}, \citenamefont
  {Plissard}, \citenamefont {Bakkers},\ and\ \citenamefont
  {Kouwenhoven}}]{Szombati2016}%
  \BibitemOpen
  \bibfield  {author} {\bibinfo {author} {\bibfnamefont {D.~B.}\ \bibnamefont
  {Szombati}}, \bibinfo {author} {\bibfnamefont {S.}~\bibnamefont
  {Nadj-Perge}}, \bibinfo {author} {\bibfnamefont {D.}~\bibnamefont {Car}},
  \bibinfo {author} {\bibfnamefont {S.~R.}\ \bibnamefont {Plissard}}, \bibinfo
  {author} {\bibfnamefont {E.~P. A.~M.}\ \bibnamefont {Bakkers}},\ and\
  \bibinfo {author} {\bibfnamefont {L.~P.}\ \bibnamefont {Kouwenhoven}},\
  }\bibfield  {title} {\bibinfo {title} {{Josephson $\varphi_0$-junction in
  nanowire quantum dots}},\ }\href {https://doi.org/10.1038/nphys3742}
  {\bibfield  {journal} {\bibinfo  {journal} {Nature Physics}\ }\textbf
  {\bibinfo {volume} {12}},\ \bibinfo {pages} {568} (\bibinfo {year}
  {2016})}\BibitemShut {NoStop}%
\bibitem [{\citenamefont {Assouline}\ \emph {et~al.}(2019)\citenamefont
  {Assouline}, \citenamefont {Feuillet-Palma}, \citenamefont {Bergeal},
  \citenamefont {Zhang}, \citenamefont {Mottaghizadeh}, \citenamefont
  {Zimmers}, \citenamefont {Lhuillier}, \citenamefont {Eddrie}, \citenamefont
  {Atkinson}, \citenamefont {Aprili},\ and\ \citenamefont
  {Aubin}}]{Assouline2019}%
  \BibitemOpen
  \bibfield  {author} {\bibinfo {author} {\bibfnamefont {A.}~\bibnamefont
  {Assouline}}, \bibinfo {author} {\bibfnamefont {C.}~\bibnamefont
  {Feuillet-Palma}}, \bibinfo {author} {\bibfnamefont {N.}~\bibnamefont
  {Bergeal}}, \bibinfo {author} {\bibfnamefont {T.}~\bibnamefont {Zhang}},
  \bibinfo {author} {\bibfnamefont {A.}~\bibnamefont {Mottaghizadeh}}, \bibinfo
  {author} {\bibfnamefont {A.}~\bibnamefont {Zimmers}}, \bibinfo {author}
  {\bibfnamefont {E.}~\bibnamefont {Lhuillier}}, \bibinfo {author}
  {\bibfnamefont {M.}~\bibnamefont {Eddrie}}, \bibinfo {author} {\bibfnamefont
  {P.}~\bibnamefont {Atkinson}}, \bibinfo {author} {\bibfnamefont
  {M.}~\bibnamefont {Aprili}},\ and\ \bibinfo {author} {\bibfnamefont
  {H.}~\bibnamefont {Aubin}},\ }\bibfield  {title} {\bibinfo {title}
  {{Spin-Orbit induced phase-shift in Bi2Se3 Josephson junctions}},\ }\href
  {https://doi.org/10.1038/s41467-018-08022-y} {\bibfield  {journal} {\bibinfo
  {journal} {Nature Communications}\ }\textbf {\bibinfo {volume} {10}},\
  \bibinfo {pages} {126} (\bibinfo {year} {2019})}\BibitemShut {NoStop}%
\bibitem [{\citenamefont {Mayer}\ \emph
  {et~al.}(2020{\natexlab{a}})\citenamefont {Mayer}, \citenamefont {Dartiailh},
  \citenamefont {Yuan}, \citenamefont {Wickramasinghe}, \citenamefont {Rossi},\
  and\ \citenamefont {Shabani}}]{Mayer2020b}%
  \BibitemOpen
  \bibfield  {author} {\bibinfo {author} {\bibfnamefont {W.}~\bibnamefont
  {Mayer}}, \bibinfo {author} {\bibfnamefont {M.~C.}\ \bibnamefont
  {Dartiailh}}, \bibinfo {author} {\bibfnamefont {J.}~\bibnamefont {Yuan}},
  \bibinfo {author} {\bibfnamefont {K.~S.}\ \bibnamefont {Wickramasinghe}},
  \bibinfo {author} {\bibfnamefont {E.}~\bibnamefont {Rossi}},\ and\ \bibinfo
  {author} {\bibfnamefont {J.}~\bibnamefont {Shabani}},\ }\bibfield  {title}
  {\bibinfo {title} {{Gate controlled anomalous phase shift in Al/InAs
  Josephson junctions}},\ }\href {https://doi.org/10.1038/s41467-019-14094-1}
  {\bibfield  {journal} {\bibinfo  {journal} {Nature Communications}\ }\textbf
  {\bibinfo {volume} {11}},\ \bibinfo {pages} {212} (\bibinfo {year}
  {2020}{\natexlab{a}})}\BibitemShut {NoStop}%
\bibitem [{\citenamefont {Ando}\ \emph {et~al.}(2020)\citenamefont {Ando},
  \citenamefont {Miyasaka}, \citenamefont {Li}, \citenamefont {Ishizuka},
  \citenamefont {Arakawa}, \citenamefont {Shiota}, \citenamefont {Moriyama},
  \citenamefont {Yanase},\ and\ \citenamefont {Ono}}]{Ando2020}%
  \BibitemOpen
  \bibfield  {author} {\bibinfo {author} {\bibfnamefont {F.}~\bibnamefont
  {Ando}}, \bibinfo {author} {\bibfnamefont {Y.}~\bibnamefont {Miyasaka}},
  \bibinfo {author} {\bibfnamefont {T.}~\bibnamefont {Li}}, \bibinfo {author}
  {\bibfnamefont {J.}~\bibnamefont {Ishizuka}}, \bibinfo {author}
  {\bibfnamefont {T.}~\bibnamefont {Arakawa}}, \bibinfo {author} {\bibfnamefont
  {Y.}~\bibnamefont {Shiota}}, \bibinfo {author} {\bibfnamefont
  {T.}~\bibnamefont {Moriyama}}, \bibinfo {author} {\bibfnamefont
  {Y.}~\bibnamefont {Yanase}},\ and\ \bibinfo {author} {\bibfnamefont
  {T.}~\bibnamefont {Ono}},\ }\bibfield  {title} {\bibinfo {title} {Observation
  of superconducting diode effect},\ }\href
  {https://doi.org/10.1038/s41586-020-2590-4} {\bibfield  {journal} {\bibinfo
  {journal} {Nature}\ }\textbf {\bibinfo {volume} {584}},\ \bibinfo {pages}
  {373} (\bibinfo {year} {2020})}\BibitemShut {NoStop}%
\bibitem [{Note1()}]{Note1}%
  \BibitemOpen
  \bibinfo {note} {See Supplemental Material for further
  information.}\BibitemShut {Stop}%
\bibitem [{\citenamefont {Groth}\ \emph {et~al.}(2014)\citenamefont {Groth},
  \citenamefont {Wimmer}, \citenamefont {Akhmerov},\ and\ \citenamefont
  {Waintal}}]{Groth2014}%
  \BibitemOpen
  \bibfield  {author} {\bibinfo {author} {\bibfnamefont {C.~W.}\ \bibnamefont
  {Groth}}, \bibinfo {author} {\bibfnamefont {M.}~\bibnamefont {Wimmer}},
  \bibinfo {author} {\bibfnamefont {A.~R.}\ \bibnamefont {Akhmerov}},\ and\
  \bibinfo {author} {\bibfnamefont {X.}~\bibnamefont {Waintal}},\ }\bibfield
  {title} {\bibinfo {title} {{Kwant: a software package for quantum
  transport}},\ }\href {https://doi.org/10.1088/1367-2630/16/6/063065}
  {\bibfield  {journal} {\bibinfo  {journal} {New J. Phys.}\ }\textbf {\bibinfo
  {volume} {16}},\ \bibinfo {pages} {063065} (\bibinfo {year}
  {2014})}\BibitemShut {NoStop}%
\bibitem [{\citenamefont {Mayer}\ \emph
  {et~al.}(2020{\natexlab{b}})\citenamefont {Mayer}, \citenamefont {Schiela},
  \citenamefont {Yuan}, \citenamefont {Hatefipour}, \citenamefont {Sarney},
  \citenamefont {Svensson}, \citenamefont {Leff}, \citenamefont {Campos},
  \citenamefont {Wickramasinghe}, \citenamefont {Dartiailh}, \citenamefont
  {{\v{Z}}uti{\'{c}}},\ and\ \citenamefont {Shabani}}]{Mayer2020}%
  \BibitemOpen
  \bibfield  {author} {\bibinfo {author} {\bibfnamefont {W.}~\bibnamefont
  {Mayer}}, \bibinfo {author} {\bibfnamefont {W.~F.}\ \bibnamefont {Schiela}},
  \bibinfo {author} {\bibfnamefont {J.}~\bibnamefont {Yuan}}, \bibinfo {author}
  {\bibfnamefont {M.}~\bibnamefont {Hatefipour}}, \bibinfo {author}
  {\bibfnamefont {W.~L.}\ \bibnamefont {Sarney}}, \bibinfo {author}
  {\bibfnamefont {S.~P.}\ \bibnamefont {Svensson}}, \bibinfo {author}
  {\bibfnamefont {A.~C.}\ \bibnamefont {Leff}}, \bibinfo {author}
  {\bibfnamefont {T.}~\bibnamefont {Campos}}, \bibinfo {author} {\bibfnamefont
  {K.~S.}\ \bibnamefont {Wickramasinghe}}, \bibinfo {author} {\bibfnamefont
  {M.~C.}\ \bibnamefont {Dartiailh}}, \bibinfo {author} {\bibfnamefont
  {I.}~\bibnamefont {{\v{Z}}uti{\'{c}}}},\ and\ \bibinfo {author}
  {\bibfnamefont {J.}~\bibnamefont {Shabani}},\ }\bibfield  {title} {\bibinfo
  {title} {{Superconducting Proximity Effect in InAsSb Surface Quantum Wells
  with In Situ Al Contacts}},\ }\href {https://doi.org/10.1021/acsaelm.0c00269}
  {\bibfield  {journal} {\bibinfo  {journal} {ACS Appl. Electron. Mater.}\
  }\textbf {\bibinfo {volume} {2}},\ \bibinfo {pages} {2351} (\bibinfo {year}
  {2020}{\natexlab{b}})}\BibitemShut {NoStop}%
\bibitem [{\citenamefont {Vurgaftman}\ \emph {et~al.}(2001)\citenamefont
  {Vurgaftman}, \citenamefont {Meyer},\ and\ \citenamefont
  {Ram-Mohan}}]{Vurgaftman2001}%
  \BibitemOpen
  \bibfield  {author} {\bibinfo {author} {\bibfnamefont {I.}~\bibnamefont
  {Vurgaftman}}, \bibinfo {author} {\bibfnamefont {J.~R.}\ \bibnamefont
  {Meyer}},\ and\ \bibinfo {author} {\bibfnamefont {L.~R.}\ \bibnamefont
  {Ram-Mohan}},\ }\bibfield  {title} {\bibinfo {title} {{Band parameters for
  III–V compound semiconductors and their alloys}},\ }\href
  {https://doi.org/10.1063/1.1368156} {\bibfield  {journal} {\bibinfo
  {journal} {J. Appl. Phys.}\ }\textbf {\bibinfo {volume} {89}},\ \bibinfo
  {pages} {5815} (\bibinfo {year} {2001})}\BibitemShut {NoStop}%
\bibitem [{\citenamefont {Fabian}\ \emph {et~al.}(2007)\citenamefont {Fabian},
  \citenamefont {Matos-Abiague}, \citenamefont {Ertler}, \citenamefont
  {Stano},\ and\ \citenamefont {{\v{Z}}uti{\'{c}}}}]{Fabian2007}%
  \BibitemOpen
  \bibfield  {author} {\bibinfo {author} {\bibfnamefont {J.}~\bibnamefont
  {Fabian}}, \bibinfo {author} {\bibfnamefont {A.}~\bibnamefont
  {Matos-Abiague}}, \bibinfo {author} {\bibfnamefont {C.}~\bibnamefont
  {Ertler}}, \bibinfo {author} {\bibfnamefont {P.}~\bibnamefont {Stano}},\ and\
  \bibinfo {author} {\bibfnamefont {I.}~\bibnamefont {{\v{Z}}uti{\'{c}}}},\
  }\bibfield  {title} {\bibinfo {title} {{Semiconductor Spintronics}},\ }\href
  {http://www.physics.sk/aps/pub.php?y=2007&pub=aps-07-04} {\bibfield
  {journal} {\bibinfo  {journal} {Acta Phys. Slovaca}\ }\textbf {\bibinfo
  {volume} {57}},\ \bibinfo {pages} {565} (\bibinfo {year} {2007})}\BibitemShut
  {NoStop}%
\bibitem [{\citenamefont {Buzdin}(2008{\natexlab{b}})}]{Buzdin2008}%
  \BibitemOpen
  \bibfield  {author} {\bibinfo {author} {\bibfnamefont {A.}~\bibnamefont
  {Buzdin}},\ }\bibfield  {title} {\bibinfo {title} {{Direct Coupling Between
  Magnetism and Superconducting Current in the Josephson
  ${\ensuremath{\varphi}}_{0}$ Junction}},\ }\href
  {https://doi.org/10.1103/PhysRevLett.101.107005} {\bibfield  {journal}
  {\bibinfo  {journal} {Phys. Rev. Lett.}\ }\textbf {\bibinfo {volume} {101}},\
  \bibinfo {pages} {107005} (\bibinfo {year} {2008}{\natexlab{b}})}\BibitemShut
  {NoStop}%
\bibitem [{\citenamefont {Seraide}\ and\ \citenamefont
  {Hai}(2002)}]{Seraide2002}%
  \BibitemOpen
  \bibfield  {author} {\bibinfo {author} {\bibfnamefont {R.~M.}\ \bibnamefont
  {Seraide}}\ and\ \bibinfo {author} {\bibfnamefont {G.-Q.}\ \bibnamefont
  {Hai}},\ }\bibfield  {title} {\bibinfo {title} {{Low-temperature electron
  mobility in parabolic quantum wells}},\ }\href
  {https://doi.org/10.1590/S0103-97332002000200026} {\bibfield  {journal}
  {\bibinfo  {journal} {Brazilian J. Phys.}\ }\textbf {\bibinfo {volume}
  {32}},\ \bibinfo {pages} {344} (\bibinfo {year} {2002})}\BibitemShut
  {NoStop}%
\bibitem [{\citenamefont {Suominen}\ \emph {et~al.}(2017)\citenamefont
  {Suominen}, \citenamefont {Danon}, \citenamefont {Kjaergaard}, \citenamefont
  {Flensberg}, \citenamefont {Shabani}, \citenamefont {Palmstr\o{}m},
  \citenamefont {Nichele},\ and\ \citenamefont {Marcus}}]{Suominen2017}%
  \BibitemOpen
  \bibfield  {author} {\bibinfo {author} {\bibfnamefont {H.~J.}\ \bibnamefont
  {Suominen}}, \bibinfo {author} {\bibfnamefont {J.}~\bibnamefont {Danon}},
  \bibinfo {author} {\bibfnamefont {M.}~\bibnamefont {Kjaergaard}}, \bibinfo
  {author} {\bibfnamefont {K.}~\bibnamefont {Flensberg}}, \bibinfo {author}
  {\bibfnamefont {J.}~\bibnamefont {Shabani}}, \bibinfo {author} {\bibfnamefont
  {C.~J.}\ \bibnamefont {Palmstr\o{}m}}, \bibinfo {author} {\bibfnamefont
  {F.}~\bibnamefont {Nichele}},\ and\ \bibinfo {author} {\bibfnamefont {C.~M.}\
  \bibnamefont {Marcus}},\ }\bibfield  {title} {\bibinfo {title} {{Anomalous
  Fraunhofer interference in epitaxial superconductor-semiconductor Josephson
  junctions}},\ }\href {https://doi.org/10.1103/PhysRevB.95.035307} {\bibfield
  {journal} {\bibinfo  {journal} {Phys. Rev. B}\ }\textbf {\bibinfo {volume}
  {95}},\ \bibinfo {pages} {035307} (\bibinfo {year} {2017})}\BibitemShut
  {NoStop}%
\bibitem [{\citenamefont {Guiducci}\ \emph {et~al.}(2019)\citenamefont
  {Guiducci}, \citenamefont {Carrega}, \citenamefont {Taddei}, \citenamefont
  {Biasiol}, \citenamefont {Courtois}, \citenamefont {Beltram},\ and\
  \citenamefont {Heun}}]{Guiducci2019}%
  \BibitemOpen
  \bibfield  {author} {\bibinfo {author} {\bibfnamefont {S.}~\bibnamefont
  {Guiducci}}, \bibinfo {author} {\bibfnamefont {M.}~\bibnamefont {Carrega}},
  \bibinfo {author} {\bibfnamefont {F.}~\bibnamefont {Taddei}}, \bibinfo
  {author} {\bibfnamefont {G.}~\bibnamefont {Biasiol}}, \bibinfo {author}
  {\bibfnamefont {H.}~\bibnamefont {Courtois}}, \bibinfo {author}
  {\bibfnamefont {F.}~\bibnamefont {Beltram}},\ and\ \bibinfo {author}
  {\bibfnamefont {S.}~\bibnamefont {Heun}},\ }\bibfield  {title} {\bibinfo
  {title} {{Full electrostatic control of quantum interference in an extended
  trenched Josephson junction}},\ }\href
  {https://doi.org/10.1103/PhysRevB.99.235419} {\bibfield  {journal} {\bibinfo
  {journal} {Phys. Rev. B}\ }\textbf {\bibinfo {volume} {99}},\ \bibinfo
  {pages} {235419} (\bibinfo {year} {2019})}\BibitemShut {NoStop}%
\bibitem [{\citenamefont {Ideue}\ \emph {et~al.}(2017)\citenamefont {Ideue},
  \citenamefont {Hamamoto}, \citenamefont {Koshikawa}, \citenamefont {Ezawa},
  \citenamefont {Shimizu}, \citenamefont {Kaneko}, \citenamefont {Tokura},
  \citenamefont {Nagaosa},\ and\ \citenamefont {Iwasa}}]{Ideue2017}%
  \BibitemOpen
  \bibfield  {author} {\bibinfo {author} {\bibfnamefont {T.}~\bibnamefont
  {Ideue}}, \bibinfo {author} {\bibfnamefont {K.}~\bibnamefont {Hamamoto}},
  \bibinfo {author} {\bibfnamefont {S.}~\bibnamefont {Koshikawa}}, \bibinfo
  {author} {\bibfnamefont {M.}~\bibnamefont {Ezawa}}, \bibinfo {author}
  {\bibfnamefont {S.}~\bibnamefont {Shimizu}}, \bibinfo {author} {\bibfnamefont
  {Y.}~\bibnamefont {Kaneko}}, \bibinfo {author} {\bibfnamefont
  {Y.}~\bibnamefont {Tokura}}, \bibinfo {author} {\bibfnamefont
  {N.}~\bibnamefont {Nagaosa}},\ and\ \bibinfo {author} {\bibfnamefont
  {Y.}~\bibnamefont {Iwasa}},\ }\bibfield  {title} {\bibinfo {title} {Bulk
  rectification effect in a polar semiconductor},\ }\href
  {https://doi.org/10.1038/nphys4056} {\bibfield  {journal} {\bibinfo
  {journal} {Nature Physics}\ }\textbf {\bibinfo {volume} {13}},\ \bibinfo
  {pages} {578} (\bibinfo {year} {2017})}\BibitemShut {NoStop}%
\bibitem [{\citenamefont {He}\ \emph {et~al.}(2018)\citenamefont {He},
  \citenamefont {Walker}, \citenamefont {Zhang}, \citenamefont {Bruno},
  \citenamefont {Bahramy}, \citenamefont {Lee}, \citenamefont {Ramaswamy},
  \citenamefont {Cai}, \citenamefont {Heinonen}, \citenamefont {Vignale},
  \citenamefont {Baumberger},\ and\ \citenamefont {Yang}}]{He2018}%
  \BibitemOpen
  \bibfield  {author} {\bibinfo {author} {\bibfnamefont {P.}~\bibnamefont
  {He}}, \bibinfo {author} {\bibfnamefont {S.~M.}\ \bibnamefont {Walker}},
  \bibinfo {author} {\bibfnamefont {S.~S.-L.}\ \bibnamefont {Zhang}}, \bibinfo
  {author} {\bibfnamefont {F.~Y.}\ \bibnamefont {Bruno}}, \bibinfo {author}
  {\bibfnamefont {M.~S.}\ \bibnamefont {Bahramy}}, \bibinfo {author}
  {\bibfnamefont {J.~M.}\ \bibnamefont {Lee}}, \bibinfo {author} {\bibfnamefont
  {R.}~\bibnamefont {Ramaswamy}}, \bibinfo {author} {\bibfnamefont
  {K.}~\bibnamefont {Cai}}, \bibinfo {author} {\bibfnamefont {O.}~\bibnamefont
  {Heinonen}}, \bibinfo {author} {\bibfnamefont {G.}~\bibnamefont {Vignale}},
  \bibinfo {author} {\bibfnamefont {F.}~\bibnamefont {Baumberger}},\ and\
  \bibinfo {author} {\bibfnamefont {H.}~\bibnamefont {Yang}},\ }\bibfield
  {title} {\bibinfo {title} {{Observation of Out-of-Plane Spin Texture in a
  SrTiO$_{3}(111)$ Two-Dimensional Electron Gas}},\ }\href
  {https://doi.org/10.1103/PhysRevLett.120.266802} {\bibfield  {journal}
  {\bibinfo  {journal} {Phys. Rev. Lett.}\ }\textbf {\bibinfo {volume} {120}},\
  \bibinfo {pages} {266802} (\bibinfo {year} {2018})}\BibitemShut {NoStop}%
\bibitem [{\citenamefont {Bychkov}\ and\ \citenamefont
  {Rashba}(1984{\natexlab{a}})}]{Bychkov1984}%
  \BibitemOpen
  \bibfield  {author} {\bibinfo {author} {\bibfnamefont {Y.~A.}\ \bibnamefont
  {Bychkov}}\ and\ \bibinfo {author} {\bibfnamefont {E.~I.}\ \bibnamefont
  {Rashba}},\ }\bibfield  {title} {\bibinfo {title} {{Oscillatory effects and
  the magnetic susceptibility of carriers in inversion layers}},\ }\href
  {http://stacks.iop.org/0022-3719/17/i=33/a=015?key=crossref.2f8159f54a3070499f32bac53e23f947}
  {\bibfield  {journal} {\bibinfo  {journal} {J. Phys. C}\ }\textbf {\bibinfo
  {volume} {17}},\ \bibinfo {pages} {6039} (\bibinfo {year}
  {1984}{\natexlab{a}})}\BibitemShut {NoStop}%
\bibitem [{\citenamefont {Bychkov}\ and\ \citenamefont
  {Rashba}(1984{\natexlab{b}})}]{Bychkov1984b}%
  \BibitemOpen
  \bibfield  {author} {\bibinfo {author} {\bibfnamefont {Y.~A.}\ \bibnamefont
  {Bychkov}}\ and\ \bibinfo {author} {\bibfnamefont {E.~I.}\ \bibnamefont
  {Rashba}},\ }\bibfield  {title} {\bibinfo {title} {{Properties of a 2D
  electron gas with lifted spectral degeneracy}},\ }\href@noop {} {\bibfield
  {journal} {\bibinfo  {journal} {Pis'ma Zh. Eksp. Teor. Fiz.}\ }\textbf
  {\bibinfo {volume} {39}},\ \bibinfo {pages} {66} (\bibinfo {year}
  {1984}{\natexlab{b}})}\BibitemShut {NoStop}%
\bibitem [{Byc(1984)}]{Bychkov1984c}%
  \BibitemOpen
  \href {http://www.jetpletters.ac.ru/ps/1264/article_19121.shtml} {\bibfield
  {journal} {\bibinfo  {journal} {JETP Lett.}\ }\textbf {\bibinfo {volume}
  {39}},\ \bibinfo {pages} {78} (\bibinfo {year} {1984})}\BibitemShut {NoStop}%
\bibitem [{\citenamefont {Dresselhaus}(1955)}]{Dresselhaus1955}%
  \BibitemOpen
  \bibfield  {author} {\bibinfo {author} {\bibfnamefont {G.}~\bibnamefont
  {Dresselhaus}},\ }\bibfield  {title} {\bibinfo {title} {{Spin-Orbit Coupling
  Effects in Zinc Blende Structures}},\ }\href
  {http://dx.doi.org/10.1103/PhysRev.100.580} {\bibfield  {journal} {\bibinfo
  {journal} {Phys. Rev.}\ }\textbf {\bibinfo {volume} {100}},\ \bibinfo {pages}
  {580} (\bibinfo {year} {1955})}\BibitemShut {NoStop}%
\bibitem [{\citenamefont {Calsaverini}\ \emph {et~al.}(2008)\citenamefont
  {Calsaverini}, \citenamefont {Bernardes}, \citenamefont {Egues},\ and\
  \citenamefont {Loss}}]{Calsaverini2008PRB}%
  \BibitemOpen
  \bibfield  {author} {\bibinfo {author} {\bibfnamefont {R.~S.}\ \bibnamefont
  {Calsaverini}}, \bibinfo {author} {\bibfnamefont {E.}~\bibnamefont
  {Bernardes}}, \bibinfo {author} {\bibfnamefont {J.~C.}\ \bibnamefont
  {Egues}},\ and\ \bibinfo {author} {\bibfnamefont {D.}~\bibnamefont {Loss}},\
  }\bibfield  {title} {\bibinfo {title} {Intersubband-induced spin-orbit
  interaction in quantum wells},\ }\href
  {https://doi.org/10.1103/PhysRevB.78.155313} {\bibfield  {journal} {\bibinfo
  {journal} {Phys. Rev. B}\ }\textbf {\bibinfo {volume} {78}},\ \bibinfo
  {pages} {155313} (\bibinfo {year} {2008})}\BibitemShut {NoStop}%
\bibitem [{\citenamefont {Fu}\ and\ \citenamefont {Egues}(2015)}]{Fu2015PRB}%
  \BibitemOpen
  \bibfield  {author} {\bibinfo {author} {\bibfnamefont {J.}~\bibnamefont
  {Fu}}\ and\ \bibinfo {author} {\bibfnamefont {J.~C.}\ \bibnamefont {Egues}},\
  }\bibfield  {title} {\bibinfo {title} {{Spin-orbit interaction in GaAs wells:
  From one to two subbands}},\ }\href
  {https://doi.org/10.1103/PhysRevB.91.075408} {\bibfield  {journal} {\bibinfo
  {journal} {Phys. Rev. B}\ }\textbf {\bibinfo {volume} {91}},\ \bibinfo
  {pages} {075408} (\bibinfo {year} {2015})}\BibitemShut {NoStop}%
\bibitem [{\citenamefont {Antipov}\ \emph {et~al.}(2018)\citenamefont
  {Antipov}, \citenamefont {Bargerbos}, \citenamefont {Winkler}, \citenamefont
  {Bauer}, \citenamefont {Rossi},\ and\ \citenamefont
  {Lutchyn}}]{Antipov2018PRX}%
  \BibitemOpen
  \bibfield  {author} {\bibinfo {author} {\bibfnamefont {A.~E.}\ \bibnamefont
  {Antipov}}, \bibinfo {author} {\bibfnamefont {A.}~\bibnamefont {Bargerbos}},
  \bibinfo {author} {\bibfnamefont {G.~W.}\ \bibnamefont {Winkler}}, \bibinfo
  {author} {\bibfnamefont {B.}~\bibnamefont {Bauer}}, \bibinfo {author}
  {\bibfnamefont {E.}~\bibnamefont {Rossi}},\ and\ \bibinfo {author}
  {\bibfnamefont {R.~M.}\ \bibnamefont {Lutchyn}},\ }\bibfield  {title}
  {\bibinfo {title} {{Effects of Gate-Induced Electric Fields on Semiconductor
  Majorana Nanowires}},\ }\href {https://doi.org/10.1103/PhysRevX.8.031041}
  {\bibfield  {journal} {\bibinfo  {journal} {Phys. Rev. X}\ }\textbf {\bibinfo
  {volume} {8}},\ \bibinfo {pages} {031041} (\bibinfo {year}
  {2018})}\BibitemShut {NoStop}%
\bibitem [{\citenamefont {Mikkelsen}\ \emph {et~al.}(2018)\citenamefont
  {Mikkelsen}, \citenamefont {Kotetes}, \citenamefont {Krogstrup},\ and\
  \citenamefont {Flensberg}}]{Mikkelsen2018PRX}%
  \BibitemOpen
  \bibfield  {author} {\bibinfo {author} {\bibfnamefont {A.~E.~G.}\
  \bibnamefont {Mikkelsen}}, \bibinfo {author} {\bibfnamefont {P.}~\bibnamefont
  {Kotetes}}, \bibinfo {author} {\bibfnamefont {P.}~\bibnamefont {Krogstrup}},\
  and\ \bibinfo {author} {\bibfnamefont {K.}~\bibnamefont {Flensberg}},\
  }\bibfield  {title} {\bibinfo {title} {{Hybridization at
  Superconductor-Semiconductor Interfaces}},\ }\href
  {https://doi.org/10.1103/PhysRevX.8.031040} {\bibfield  {journal} {\bibinfo
  {journal} {Phys. Rev. X}\ }\textbf {\bibinfo {volume} {8}},\ \bibinfo {pages}
  {031040} (\bibinfo {year} {2018})}\BibitemShut {NoStop}%
\bibitem [{\citenamefont {De~Gennes}(1989)}]{DeGennes1989}%
  \BibitemOpen
  \bibfield  {author} {\bibinfo {author} {\bibfnamefont {P.~G.}\ \bibnamefont
  {De~Gennes}},\ }\href@noop {} {\emph {\bibinfo {title} {{Superconductivity of
  Metals and Alloys}}}}\ (\bibinfo  {publisher} {Addison Wesley, Redwood
  City},\ \bibinfo {year} {1989})\BibitemShut {NoStop}%
\bibitem [{Note2()}]{Note2}%
  \BibitemOpen
  \bibinfo {note} {Notice that there is an evident typo in the Supplemental
  Material of Ref.~\cite {Itahashi2020}; the units for the width must be $\upmu
  $m and not mm, as confirmed by one of the authors in a private
  communication.}\BibitemShut {Stop}%
\bibitem [{\citenamefont {Ambegaokar}\ and\ \citenamefont
  {Halperin}(1969)}]{AmbegaokarHalperin1969}%
  \BibitemOpen
  \bibfield  {author} {\bibinfo {author} {\bibfnamefont {V.}~\bibnamefont
  {Ambegaokar}}\ and\ \bibinfo {author} {\bibfnamefont {B.~I.}\ \bibnamefont
  {Halperin}},\ }\bibfield  {title} {\bibinfo {title} {{Voltage Due to Thermal
  Noise in the dc Josephson Effect}},\ }\href
  {https://doi.org/10.1103/PhysRevLett.22.1364} {\bibfield  {journal} {\bibinfo
   {journal} {Phys. Rev. Lett.}\ }\textbf {\bibinfo {volume} {22}},\ \bibinfo
  {pages} {1364} (\bibinfo {year} {1969})}\BibitemShut {NoStop}%
\bibitem [{\citenamefont {Tinkham}(2004)}]{Tinkhambook}%
  \BibitemOpen
  \bibfield  {author} {\bibinfo {author} {\bibfnamefont {M.}~\bibnamefont
  {Tinkham}},\ }\href {http://www.worldcat.org/isbn/0486435032} {\emph
  {\bibinfo {title} {{Introduction to Superconductivity: Second Edition (Dover
  Books on Physics) (Vol i)}}}},\ \bibinfo {edition} {2nd}\ ed.\ (\bibinfo
  {publisher} {Dover Publications},\ \bibinfo {year} {2004})\BibitemShut
  {NoStop}%
\bibitem [{\citenamefont {Koralek}\ \emph {et~al.}(2009)\citenamefont
  {Koralek}, \citenamefont {Weber}, \citenamefont {Orenstein}, \citenamefont
  {Bernevig}, \citenamefont {Zhang}, \citenamefont {Mack},\ and\ \citenamefont
  {Awschalom}}]{Koralek2009Nat}%
  \BibitemOpen
  \bibfield  {author} {\bibinfo {author} {\bibfnamefont {J.~D.}\ \bibnamefont
  {Koralek}}, \bibinfo {author} {\bibfnamefont {C.~P.}\ \bibnamefont {Weber}},
  \bibinfo {author} {\bibfnamefont {J.}~\bibnamefont {Orenstein}}, \bibinfo
  {author} {\bibfnamefont {B.~A.}\ \bibnamefont {Bernevig}}, \bibinfo {author}
  {\bibfnamefont {S.-C.}\ \bibnamefont {Zhang}}, \bibinfo {author}
  {\bibfnamefont {S.}~\bibnamefont {Mack}},\ and\ \bibinfo {author}
  {\bibfnamefont {D.}~\bibnamefont {Awschalom}},\ }\bibfield  {title} {\bibinfo
  {title} {{Emergence of the persistent spin helix in semiconductor quantum
  wells}},\ }\href@noop {} {\bibfield  {journal} {\bibinfo  {journal} {Nature}\
  }\textbf {\bibinfo {volume} {458}},\ \bibinfo {pages} {610} (\bibinfo {year}
  {2009})}\BibitemShut {NoStop}%
\bibitem [{\citenamefont {Kane}(1957)}]{Kane1957}%
  \BibitemOpen
  \bibfield  {author} {\bibinfo {author} {\bibfnamefont {E.~O.}\ \bibnamefont
  {Kane}},\ }\bibfield  {title} {\bibinfo {title} {{Band structure of indium
  antimonide}},\ }\href@noop {} {\bibfield  {journal} {\bibinfo  {journal}
  {Journal of Physics and Chemistry of Solids}\ }\textbf {\bibinfo {volume}
  {1}},\ \bibinfo {pages} {249} (\bibinfo {year} {1957})}\BibitemShut {NoStop}%
\bibitem [{\citenamefont {Winkler}(2003)}]{Winkler2003spin}%
  \BibitemOpen
  \bibfield  {author} {\bibinfo {author} {\bibfnamefont {R.}~\bibnamefont
  {Winkler}},\ }\href@noop {} {\emph {\bibinfo {title} {{Spin-orbit Coupling
  Effects in Two-Dimensional Electron and Hole Systems}}}},\ \bibinfo {series}
  {Physics and Astronomy Online Library}\ No.\ \bibinfo {number} {no. 191}\
  (\bibinfo  {publisher} {Springer},\ \bibinfo {year} {2003})\BibitemShut
  {NoStop}%
\bibitem [{\citenamefont {Campos}\ \emph {et~al.}(2018)\citenamefont {Campos},
  \citenamefont {Faria~Junior}, \citenamefont {Gmitra}, \citenamefont
  {Sipahi},\ and\ \citenamefont {Fabian}}]{Campos2018PRB}%
  \BibitemOpen
  \bibfield  {author} {\bibinfo {author} {\bibfnamefont {T.}~\bibnamefont
  {Campos}}, \bibinfo {author} {\bibfnamefont {P.~E.}\ \bibnamefont
  {Faria~Junior}}, \bibinfo {author} {\bibfnamefont {M.}~\bibnamefont
  {Gmitra}}, \bibinfo {author} {\bibfnamefont {G.~M.}\ \bibnamefont {Sipahi}},\
  and\ \bibinfo {author} {\bibfnamefont {J.}~\bibnamefont {Fabian}},\
  }\bibfield  {title} {\bibinfo {title} {{Spin-orbit coupling effects in
  zinc-blende InSb and wurtzite InAs nanowires: Realistic calculations with
  multiband $\mathbf{k}\ifmmode\cdot\else\textperiodcentered\fi{}\mathbf{p}$
  method}},\ }\href {https://doi.org/10.1103/PhysRevB.97.245402} {\bibfield
  {journal} {\bibinfo  {journal} {Phys. Rev. B}\ }\textbf {\bibinfo {volume}
  {97}},\ \bibinfo {pages} {245402} (\bibinfo {year} {2018})}\BibitemShut
  {NoStop}%
\bibitem [{\citenamefont {Chantis}\ \emph {et~al.}(2006)\citenamefont
  {Chantis}, \citenamefont {van Schilfgaarde},\ and\ \citenamefont
  {Kotani}}]{Chantis2006PRL}%
  \BibitemOpen
  \bibfield  {author} {\bibinfo {author} {\bibfnamefont {A.~N.}\ \bibnamefont
  {Chantis}}, \bibinfo {author} {\bibfnamefont {M.}~\bibnamefont {van
  Schilfgaarde}},\ and\ \bibinfo {author} {\bibfnamefont {T.}~\bibnamefont
  {Kotani}},\ }\bibfield  {title} {\bibinfo {title} {{Ab Initio Prediction of
  Conduction Band Spin Splitting in Zinc Blende Semiconductors}},\ }\href
  {https://doi.org/10.1103/PhysRevLett.96.086405} {\bibfield  {journal}
  {\bibinfo  {journal} {Phys. Rev. Lett.}\ }\textbf {\bibinfo {volume} {96}},\
  \bibinfo {pages} {086405} (\bibinfo {year} {2006})}\BibitemShut {NoStop}%
\bibitem [{Note3()}]{Note3}%
  \BibitemOpen
  \bibinfo {note} {As a comment, in the $\protect \mathbf {k} \cdot \protect
  \mathbf {p}$ study, the $\protect \hat {z}$-axis has shifted origin and
  reversed orientation. For the~\protect \textsc {Kwant}~implementation, it is
  more convenient to use the present coordinate system. However, it is clear
  that the two different conventions lead to the same physical
  conclusions.}\BibitemShut {Stop}%
\bibitem [{\citenamefont {Shabani}\ \emph {et~al.}(2016)\citenamefont
  {Shabani}, \citenamefont {Kjaergaard}, \citenamefont {Suominen},
  \citenamefont {Kim}, \citenamefont {Nichele}, \citenamefont {Pakrouski},
  \citenamefont {Stankevic}, \citenamefont {Lutchyn}, \citenamefont
  {Krogstrup}, \citenamefont {Feidenhans'l}, \citenamefont {Kraemer},
  \citenamefont {Nayak}, \citenamefont {Troyer}, \citenamefont {Marcus},\ and\
  \citenamefont {Palmstr{\o}m}}]{Shabani2016}%
  \BibitemOpen
  \bibfield  {author} {\bibinfo {author} {\bibfnamefont {J.}~\bibnamefont
  {Shabani}}, \bibinfo {author} {\bibfnamefont {M.}~\bibnamefont {Kjaergaard}},
  \bibinfo {author} {\bibfnamefont {H.~J.}\ \bibnamefont {Suominen}}, \bibinfo
  {author} {\bibfnamefont {Y.}~\bibnamefont {Kim}}, \bibinfo {author}
  {\bibfnamefont {F.}~\bibnamefont {Nichele}}, \bibinfo {author} {\bibfnamefont
  {K.}~\bibnamefont {Pakrouski}}, \bibinfo {author} {\bibfnamefont
  {T.}~\bibnamefont {Stankevic}}, \bibinfo {author} {\bibfnamefont {R.~M.}\
  \bibnamefont {Lutchyn}}, \bibinfo {author} {\bibfnamefont {P.}~\bibnamefont
  {Krogstrup}}, \bibinfo {author} {\bibfnamefont {R.}~\bibnamefont
  {Feidenhans'l}}, \bibinfo {author} {\bibfnamefont {S.}~\bibnamefont
  {Kraemer}}, \bibinfo {author} {\bibfnamefont {C.}~\bibnamefont {Nayak}},
  \bibinfo {author} {\bibfnamefont {M.}~\bibnamefont {Troyer}}, \bibinfo
  {author} {\bibfnamefont {C.~M.}\ \bibnamefont {Marcus}},\ and\ \bibinfo
  {author} {\bibfnamefont {C.~J.}\ \bibnamefont {Palmstr{\o}m}},\ }\bibfield
  {title} {\bibinfo {title} {{Two-dimensional epitaxial
  superconductor-semiconductor heterostructures: A platform for topological
  superconducting networks}},\ }\href
  {https://doi.org/10.1103/PhysRevB.93.155402} {\bibfield  {journal} {\bibinfo
  {journal} {Phys. Rev. B}\ }\textbf {\bibinfo {volume} {93}},\ \bibinfo
  {pages} {155402} (\bibinfo {year} {2016})}\BibitemShut {NoStop}%
\bibitem [{\citenamefont {Alidoust}\ \emph {et~al.}(2021)\citenamefont
  {Alidoust}, \citenamefont {Shen},\ and\ \citenamefont {\ifmmode
  \check{Z}\else \v{Z}\fi{}uti\ifmmode~\acute{c}\else
  \'{c}\fi{}}}]{Alidoust2021}%
  \BibitemOpen
  \bibfield  {author} {\bibinfo {author} {\bibfnamefont {M.}~\bibnamefont
  {Alidoust}}, \bibinfo {author} {\bibfnamefont {C.}~\bibnamefont {Shen}},\
  and\ \bibinfo {author} {\bibfnamefont {I.}~\bibnamefont {\ifmmode
  \check{Z}\else \v{Z}\fi{}uti\ifmmode~\acute{c}\else \'{c}\fi{}}},\ }\bibfield
   {title} {\bibinfo {title} {{Cubic spin-orbit coupling and anomalous
  Josephson effect in planar junctions}},\ }\href
  {https://doi.org/10.1103/PhysRevB.103.L060503} {\bibfield  {journal}
  {\bibinfo  {journal} {Phys. Rev. B}\ }\textbf {\bibinfo {volume} {103}},\
  \bibinfo {pages} {L060503} (\bibinfo {year} {2021})}\BibitemShut {NoStop}%
\bibitem [{Note4()}]{Note4}%
  \BibitemOpen
  \bibinfo {note} {Note that we corrected a typo in~$ V_\protect \mathrm
  {conf}(X) $ that has been present in~Ref.~\cite {Seraide2002}.}\BibitemShut
  {Stop}%
\bibitem [{\citenamefont {Ostroukh}\ \emph {et~al.}(2016)\citenamefont
  {Ostroukh}, \citenamefont {Baxevanis}, \citenamefont {Akhmerov},\ and\
  \citenamefont {Beenakker}}]{Ostroukh2016}%
  \BibitemOpen
  \bibfield  {author} {\bibinfo {author} {\bibfnamefont {V.~P.}\ \bibnamefont
  {Ostroukh}}, \bibinfo {author} {\bibfnamefont {B.}~\bibnamefont {Baxevanis}},
  \bibinfo {author} {\bibfnamefont {A.~R.}\ \bibnamefont {Akhmerov}},\ and\
  \bibinfo {author} {\bibfnamefont {C.~W.~J.}\ \bibnamefont {Beenakker}},\
  }\bibfield  {title} {\bibinfo {title} {{Two-dimensional Josephson vortex
  lattice and anomalously slow decay of the Fraunhofer oscillations in a
  ballistic SNS junction with a warped Fermi surface}},\ }\href
  {https://doi.org/10.1103/PhysRevB.94.094514} {\bibfield  {journal} {\bibinfo
  {journal} {Phys. Rev. B}\ }\textbf {\bibinfo {volume} {94}},\ \bibinfo
  {pages} {094514} (\bibinfo {year} {2016})}\BibitemShut {NoStop}%
\bibitem [{\citenamefont {Zuo}\ \emph {et~al.}(2017)\citenamefont {Zuo},
  \citenamefont {Mourik}, \citenamefont {Szombati}, \citenamefont {Nijholt},
  \citenamefont {van Woerkom}, \citenamefont {Geresdi}, \citenamefont {Chen},
  \citenamefont {Ostroukh}, \citenamefont {Akhmerov}, \citenamefont {Plissard},
  \citenamefont {Car}, \citenamefont {Bakkers}, \citenamefont {Pikulin},
  \citenamefont {Kouwenhoven},\ and\ \citenamefont {Frolov}}]{Zuo2017}%
  \BibitemOpen
  \bibfield  {author} {\bibinfo {author} {\bibfnamefont {K.}~\bibnamefont
  {Zuo}}, \bibinfo {author} {\bibfnamefont {V.}~\bibnamefont {Mourik}},
  \bibinfo {author} {\bibfnamefont {D.~B.}\ \bibnamefont {Szombati}}, \bibinfo
  {author} {\bibfnamefont {B.}~\bibnamefont {Nijholt}}, \bibinfo {author}
  {\bibfnamefont {D.~J.}\ \bibnamefont {van Woerkom}}, \bibinfo {author}
  {\bibfnamefont {A.}~\bibnamefont {Geresdi}}, \bibinfo {author} {\bibfnamefont
  {J.}~\bibnamefont {Chen}}, \bibinfo {author} {\bibfnamefont {V.~P.}\
  \bibnamefont {Ostroukh}}, \bibinfo {author} {\bibfnamefont {A.~R.}\
  \bibnamefont {Akhmerov}}, \bibinfo {author} {\bibfnamefont {S.~R.}\
  \bibnamefont {Plissard}}, \bibinfo {author} {\bibfnamefont {D.}~\bibnamefont
  {Car}}, \bibinfo {author} {\bibfnamefont {E.~P.}\ \bibnamefont {Bakkers}},
  \bibinfo {author} {\bibfnamefont {D.~I.}\ \bibnamefont {Pikulin}}, \bibinfo
  {author} {\bibfnamefont {L.~P.}\ \bibnamefont {Kouwenhoven}},\ and\ \bibinfo
  {author} {\bibfnamefont {S.~M.}\ \bibnamefont {Frolov}},\ }\bibfield  {title}
  {\bibinfo {title} {{Supercurrent Interference in Few-Mode Nanowire Josephson
  Junctions}},\ }\href {https://doi.org/10.1103/PhysRevLett.119.187704}
  {\bibfield  {journal} {\bibinfo  {journal} {Phys. Rev. Lett.}\ }\textbf
  {\bibinfo {volume} {119}},\ \bibinfo {pages} {187704} (\bibinfo {year}
  {2017})}\BibitemShut {NoStop}%
\bibitem [{\citenamefont {Blonder}\ \emph {et~al.}(1982)\citenamefont
  {Blonder}, \citenamefont {Tinkham},\ and\ \citenamefont
  {Klapwijk}}]{Blonder1982}%
  \BibitemOpen
  \bibfield  {author} {\bibinfo {author} {\bibfnamefont {G.~E.}\ \bibnamefont
  {Blonder}}, \bibinfo {author} {\bibfnamefont {M.}~\bibnamefont {Tinkham}},\
  and\ \bibinfo {author} {\bibfnamefont {T.~M.}\ \bibnamefont {Klapwijk}},\
  }\bibfield  {title} {\bibinfo {title} {{Transition from metallic to tunneling
  regimes in superconducting microconstrictions: Excess current, charge
  imbalance, and supercurrent conversion}},\ }\href
  {https://doi.org/10.1103/PhysRevB.25.4515} {\bibfield  {journal} {\bibinfo
  {journal} {Phys. Rev. B}\ }\textbf {\bibinfo {volume} {25}},\ \bibinfo
  {pages} {4515} (\bibinfo {year} {1982})}\BibitemShut {NoStop}%
\bibitem [{\citenamefont {H{\"{o}}gl}\ \emph {et~al.}(2015)\citenamefont
  {H{\"{o}}gl}, \citenamefont {Matos-Abiague}, \citenamefont
  {{\v{Z}}uti{\'{c}}},\ and\ \citenamefont {Fabian}}]{Hoegl2015}%
  \BibitemOpen
  \bibfield  {author} {\bibinfo {author} {\bibfnamefont {P.}~\bibnamefont
  {H{\"{o}}gl}}, \bibinfo {author} {\bibfnamefont {A.}~\bibnamefont
  {Matos-Abiague}}, \bibinfo {author} {\bibfnamefont {I.}~\bibnamefont
  {{\v{Z}}uti{\'{c}}}},\ and\ \bibinfo {author} {\bibfnamefont
  {J.}~\bibnamefont {Fabian}},\ }\bibfield  {title} {\bibinfo {title}
  {{Magnetoanisotropic Andreev Reflection in Ferromagnet-Superconductor
  Junctions}},\ }\href
  {http://journals.aps.org/prl/abstract/10.1103/PhysRevLett.115.116601}
  {\bibfield  {journal} {\bibinfo  {journal} {Phys. Rev. Lett.}\ }\textbf
  {\bibinfo {volume} {115}},\ \bibinfo {pages} {116601} (\bibinfo {year}
  {2015})}\BibitemShut {NoStop}%
\bibitem [{Hoe(2015)}]{Hoegl2015a}%
  \BibitemOpen
  \href {https://doi.org/10.1103/PhysRevLett.115.159902} {\bibfield  {journal}
  {\bibinfo  {journal} {Phys. Rev. Lett.}\ }\textbf {\bibinfo {volume} {115}},\
  \bibinfo {pages} {159902(E)} (\bibinfo {year} {2015})}\BibitemShut {NoStop}%
\bibitem [{\citenamefont {Costa}\ \emph {et~al.}(2017)\citenamefont {Costa},
  \citenamefont {H\"ogl},\ and\ \citenamefont {Fabian}}]{Costa2017}%
  \BibitemOpen
  \bibfield  {author} {\bibinfo {author} {\bibfnamefont {A.}~\bibnamefont
  {Costa}}, \bibinfo {author} {\bibfnamefont {P.}~\bibnamefont {H\"ogl}},\ and\
  \bibinfo {author} {\bibfnamefont {J.}~\bibnamefont {Fabian}},\ }\bibfield
  {title} {\bibinfo {title} {{Magnetoanisotropic Josephson effect due to
  interfacial spin-orbit fields in superconductor/ferromagnet/superconductor
  junctions}},\ }\href {https://doi.org/10.1103/PhysRevB.95.024514} {\bibfield
  {journal} {\bibinfo  {journal} {Phys. Rev. B}\ }\textbf {\bibinfo {volume}
  {95}},\ \bibinfo {pages} {024514} (\bibinfo {year} {2017})}\BibitemShut
  {NoStop}%
\end{thebibliography}%

\clearpage
\newpage

\onecolumngrid
\begin{center}
\textbf{\large Supplemental Material: A Josephson junction supercurrent diode}
\end{center}

\twocolumngrid
\beginsupplement

  \section{Materials and Methods}
    \subsection{Experiment}
    \textit{Wafer growth and initial characterization}:
    The hybrid heterostructure is epitaxially grown on an insulating InP substrate. The layer sequence features 100~nm In$_{0.52}$Al$_{0.48}$As matched buffer, 900~nm In$_{0.52}$Al$_{0.48}$As to In$_{0.84}$Al$_{0.16}$As graded buffer (18x50~nm steps), a reversed 33~nm graded buffer from In$_{0.84}$Al$_{0.16}$As to In$_{0.81}$Al$_{0.19}$As, a 25~nm In$_{0.81}$Al$_{0.19}$As layer, a 4~nm thick In$_{0.81}$Ga$_{0.19}$As bottom barrier, a 7~nm InAs quantum well, a 10~nm In$_{0.8}$Ga$_{0.2}$As top barrier, two monolayers GaAs and, finally, 7~nm aluminum film as the superconductor. 
    
    The quantum well of this wafer was patterned into a top-gated Hall-bar geometry. The aluminum was selectively removed and a Ti/Au gate electrode was deposited on top of a 40~nm aluminum-oxide layer. For this structure, the electron mobility was measured to be 22000~cm$^2$/Vs at density $ n = 0.5\cdot 10^{12}$~cm$^{-2}$ with a mean-free-path length $\ell_e \approx$ 270~nm at a gate voltage~$ V_\mathrm{g} = -1.8 \, \mathrm{V} $. Tunneling spectroscopy, performed on the same wafer via gate-defined quantum point contacts, revealed an induced gap $ \Delta^* \approx 130 \, \upmu \mathrm{eV} $ underneath the epitaxial Al~film~\cite{baumgartner2020}.

    \textit{Device fabrication}:
    All samples were fabricated using standard electron-beam lithography techniques defining first the mesa and, in a second step, the Josephson junctions. A standard wet-etching solution~(orthophosphoric acid : citric acid : hydrogen peroxide : distilled water = 1.2 : 22 : 2 : 88) was used to fabricate a well-defined mesa. The junctions were patterned by selective wet-etching of aluminum by using the etchant type~D from Transene Company. The remaining aluminum islands have a length of 1~$ \upmu \mathrm{m} $, a width of 3.15~$ \upmu \mathrm{m} $, and are separated by 100~nm. A global top gate was added by covering the whole array with 40~nm aluminum-oxide and 5~nm Ti/120~nm Au by atomic-layer deposition and electron-beam evaporation.

     \begin{figure*}[t]
        \centering
        \includegraphics[width=\textwidth]{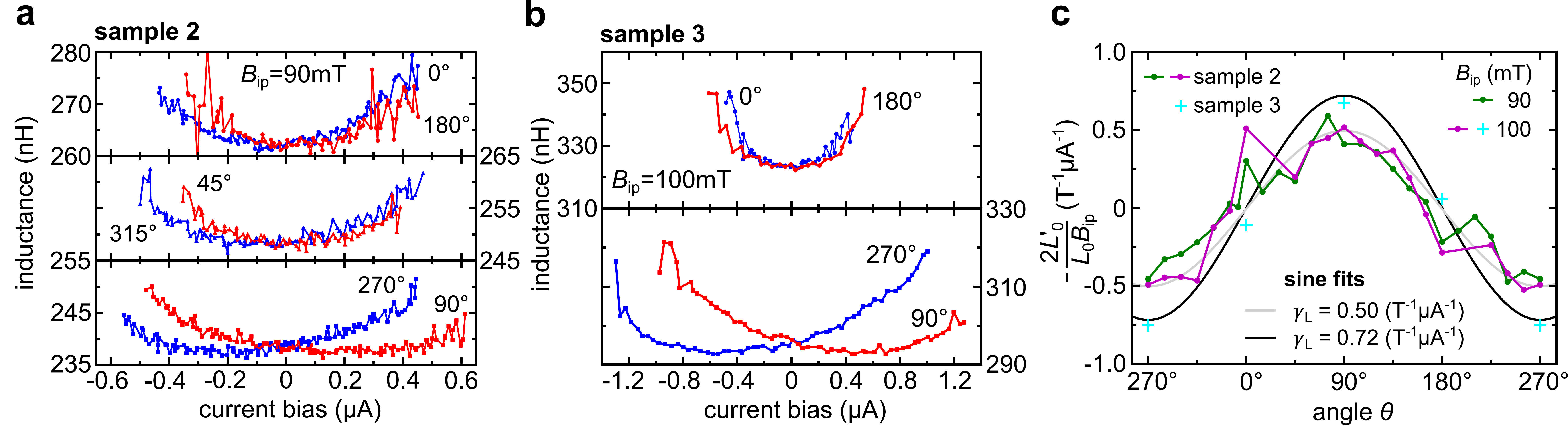}
        \caption{\textbf{a},~Current dependence of the inductance of sample~2. The junctions of this device are slightly wider, $d\simeq 165\,$nm, with a variation of width along the array of $\simeq$ 15\%. Hence, the average transparency is lower ($\bar{\tau}\simeq 0.75$) and the device is not in the ballistic regime, as opposed to sample~3 with $d=80\,$nm~(see~\textbf{b}). 
        \textbf{c},~The magnetochiral anisotropy of the former sample follows the theoretical curve for a smaller theoretical confinement-potential value in comparison with sample 3. Here, $ \gamma_L $ reaches the same values as shown in the main text.
        }
        \label{fig:other samples}
    \end{figure*}

    \textit{Measurements}:
    Our measurements were performed in a dilution refrigerator. We determine the Josephson inductance by using a cold RLC resonator mounted close to the sample (for a detailed description see Ref.~\cite{baumgartner2020}).
     The device is embedded in the resonator next to a copper coil ($ L_0 = 382$~nH), parallel to a capacitor ($ C_0 = 4$~nF), and is connected to a cold ground. The whole circuit is installed on a piezo rotator with the rotation axis perpendicular to the main magnetic field. Home-made compensation coils allow for a careful nulling of the magnetic-field component perpendicular to the sample plane. 1~k$\Omega $ resistors effectively decouple the resonant circuit from the external electrical environment. The center frequency $f \equiv (2\pi \sqrt{L_TC_0})^{-1} $ of the resonance peak is used to extract the sample inductance and is given by the capacitance $C_0$ and the total inductance $L_T$, which is the sum of the external inductance $L_0$ and the sample inductance $L$. The capacitor $C_0$ also eliminates DC electric current between the cold ground and the source and voltage contacts, which allows for complementary DC transport measurements in the same cooldown.
    The resonance frequency was chosen about 4~MHz within the range of our digital lock-in with a maximum frequency of 5~MHz. 
    The $Q$-factor is given by $Q = R_S^{-1}\sqrt{L_T/C_0}$ and is about 30. The damping resistance of the RLC circuit is typically of the order of $0.3$~$\Omega$. Already at sample resistances of a few ohms, the $Q$-factor drops towards 1 and the resonance breaks down. Since individual junctions have a normal-state resistance of $~60\,\Omega$, a single defect within a weaker junction in our one-dimensional Josephson junction array limits the maximum bias current that we can apply before the resonance breaks down.

    \subsection{Theory}
    
    \textit{Band structure}: 
    To deduce realistic values for the strength of the Bychkov--Rashba~\cite{Bychkov1984,Bychkov1984b,*Bychkov1984c} and Dresselhaus~\cite{Dresselhaus1955} spin-orbit~couplings arising inside the investigated InAs~quantum~well, and estimate their relative importance, we performed self-consistent $ \mathbf{k} \cdot \mathbf{p} $ calculations. 
    More specifically, we solved the Schr\"{o}dinger--Poisson~equation for conduction-band electrons within the quantum well~\cite{Calsaverini2008PRB,Fu2015PRB,Antipov2018PRX,Mikkelsen2018PRX}, and accounted thereby for linear and linearized cubic spin-orbit~coupling terms. 
    Our calculations give direct access to the spatial variations of the electrostatic potential inside the well and the related spin-orbit coupling parameters that can be used as an input for our actual transport simulation.

    \textit{Transport simulations}: 
    Our transport simulations are based on the Python~transport~package~\textsc{Kwant}~\cite{Groth2014}. 
    Each InAs Josephson~junction of the experimental array is described by the well-established Bogoljubov--de~Gennes~Hamiltonian~\cite{DeGennes1989}, which was discretized on a generic three-dimensional tight-binding grid as inherent to \textsc{Kwant}'s specifics. 
    Afterwards, the system was completed by adding a fictitious vertical lead, which effectively acts as self-energy from the spectral point of view. 
    \textsc{Kwant}'s functionalities allowed us to invert the Bogoljubov--de~Gennes matrix~Hamiltonian to obtain the related Green's~function and compute the Josephson~current within the aforementioned self-energy lead. 
    Following this procedure, we can derive the current--phase~relation $I(\varphi)$ of a single junction and the corresponding Josephson~inductance 
    $L(\varphi)= (\Phi_0/2\pi)[\partial I(\varphi)/\partial \varphi]^{-1}$, where $ \Phi_0=h/(2e) $ denotes the magnetic flux~quantum.
    The inductance of the whole array is given by the single-junction inductance multiplied by the number of junctions in series. 
    Eliminating the phase difference $\varphi$ from $I(\varphi)$ and $L(\varphi)$, we can compute the nonlinear inductance $L(I)$ and extract the supercurrent magnetochiral~anisotropy parameter $\gamma_L$, as described in the main text. 
    To obtain semi-quantitative simulations we used realistic material~parameters in combination with $ \mathbf{k} \cdot \mathbf{p} $~calculations tailored to the experimental setup---further details and references are provided below.

      \section{Further experimental information} 
   
   \textit{Other devices}:
   We have fabricated and measured several devices. All of them show similar phenomenology as sample~1~(described in the main text). Two other one-dimensional Josephson junction arrays are briefly discussed here. The current direction of both devices runs along the $ [1\overline{1}0] $~InAs~growth~direction.

   Sample~2 consists of 1500 Josephson junctions in series. The junction length is not uniform along the mesa: owing to imperfect lithography, the gaps separating Al islands (where Al has been selectively etched) range from 130~nm to 180~nm. Therefore, the weak links of the Josephson junctions in the array are not homogeneous in length for this sample. The Al islands are  1.05~$ \upmu \mathrm{m} $ long and 3.2~$ \upmu \mathrm{m} $ wide; their average transmission coefficient $\overline{\tau}$ is 0.75. 
   
   Sample~3 is similar to sample~1. It is made of 2250 homogeneous junctions with an etched gap of 80~nm and an average transmission coefficient $\overline{\tau} = 0.93 $. The Al~islands' length and width are 1~$ \upmu \mathrm{m} $ and 3.3 $ \upmu \mathrm{m} $, respectively. 
   
   Figures~\ref{fig:other samples}\textbf{a}--\textbf{b} show the $ L(I) $-dependencies for samples~2 and 3, at applied in-plane magnetic fields of 90~mT and 100~mT, respectively. The quality of our $ L(I) $-measurements on sample~2 is already lower at its ground state when compared to the other devices, and therefore the signal becomes too weak at finite in-plane fields. At about 100~mT, a proper evaluation gets hence challenging. Thus, we characterized this device once again at 90~mT. The inductance~asymmetry reaches a maximum for fields pointing along $ \pm \hat{y} $~($ \theta = 90^{\circ}$ and 270$^{\circ}$) and a minimum for fields along $ \pm \hat{x} $~($ \theta = 0^{\circ}$ and 180$^{\circ}$). 
    
    In Fig.~\ref{fig:other samples}\textbf{c}, we display the supercurrent magnetochiral anisotropy for both devices. Device~2 shows slightly smaller $\gamma_{L}$~(roughly $0.5 \cdot 10^6 $~T$^{-1}$A$^{-1}$), while we obtain $\gamma_{L} = 0.75 \cdot 10^6 $~T$^{-1}$A$^{-1}$ from our measurements on device~3.

    \textit{Comparison with dissipative magneto-electric effects in other materials: }\\
    In the main text, we point out that the definition of $ \gamma_S $ contains the current in the denominator---analog to Eq.~1 of the main text ($\gamma_S$ should not be confused with the magnetochiral coefficient $ \gamma_L $ for the inductance, which is the novel quantity introduced in the present work).
    Hence, it is the product $ w\gamma_S$ between the sample width $w$ and the magnetochiral coefficient $\gamma_S$ that is geometry-independent. In addition, $ w\gamma_S$ is proportional to the strength $ r_t $ of the mixing of singlet and triplet pairing amplitudes~(see Eq.~11 in the Supplemental Material of Ref.~\cite{Itahashi2020}).

    In our case, we find $ \gamma_S = 4.1 \cdot 10^6 $~T$^{-1}$A$^{-1}$, $ w = 3.15$~$\upmu$m, and thus $ w\gamma_S  = 12.9 $~T$^{-1}$A$^{-1}$m. This is of the same order of magnitude as the value $ w\gamma_S =3.2$~T$^{-1}$A$^{-1}$m that results from~\footnote{Notice that there is an evident typo in the Supplemental Material of Ref.~\cite{Itahashi2020}; the units for the width must be $\upmu$m and not mm, as confirmed by one of the authors in a private communication. } $\gamma_S=4.0\cdot10^4\,$T$^{-1}$A$^{-1}$ and $w =80$~$\upmu$m for a gate-induced surface electron system in SrTiO$_3$ reported by Itahashi \textit{et al.}~in Ref.~\cite{Itahashi2020}. 
    Hence, we conclude that the magnitude of $ w\gamma_S$ varies surprisingly little even when comparing very different materials. \\[3mm]

     \begin{figure}[t]
        \centering
        \includegraphics[width=0.475\textwidth]{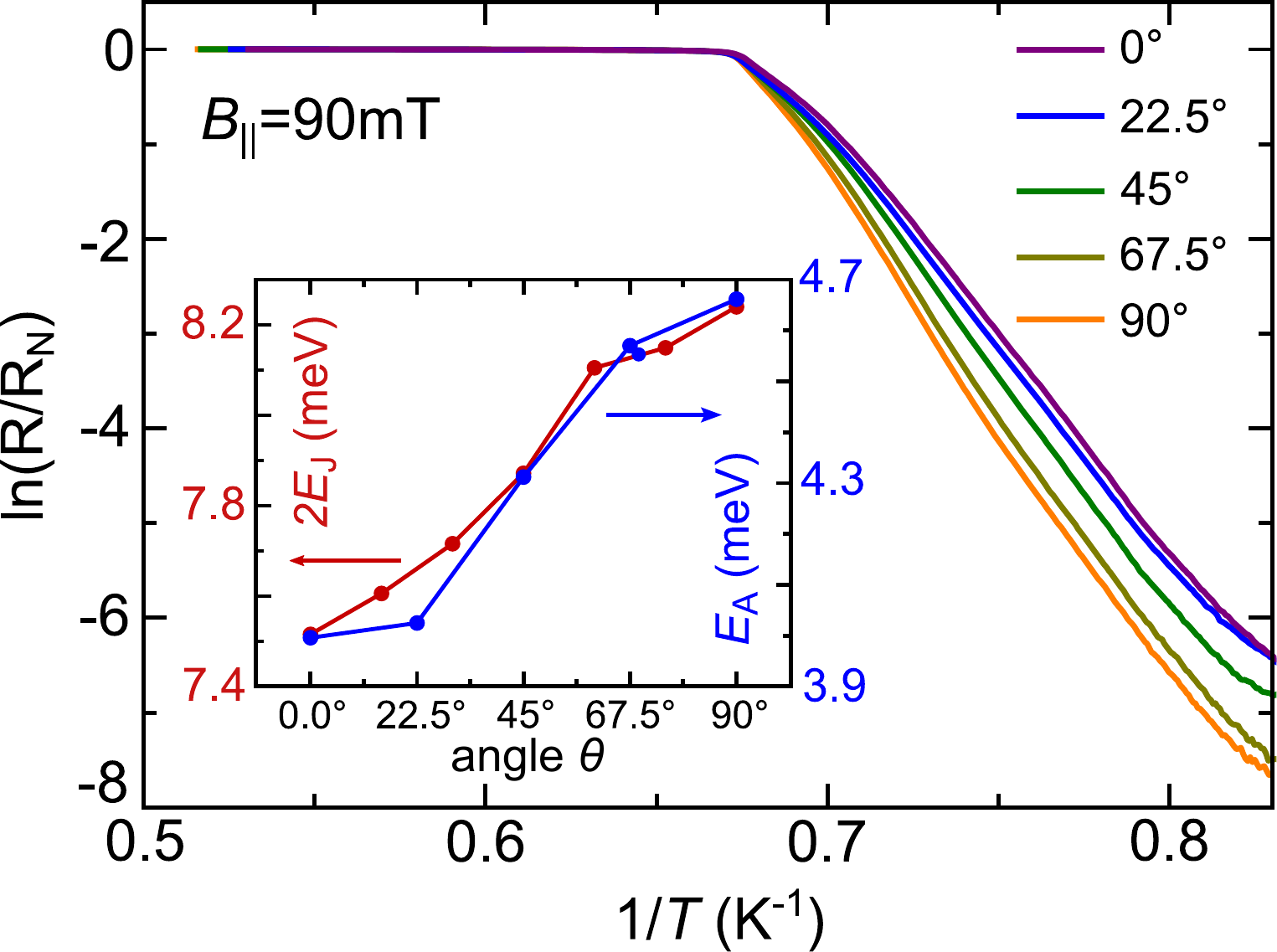}
        \caption{
        Arrhenius plots of resistance vs.~inverse temperature in an in-plane magnetic field of 90~mT for different angles $\theta$ between the current and the magnetic field ($\theta=0^{\circ}$ corresponds to the field being parallel to the current). We extract the thermal activation energy $E_\mathrm{A}(\theta) $ from the slopes in the exponential regime. 
        Inset: Activation energy $E_\mathrm{A}$ from the Arrhenius law together with (twice) the Josephson coupling energy $E_\mathrm{J}$ (see text). The latter was calculated using $L_\mathrm{0}$ in Tab.~\ref{tab:activation energy}.
        }
        \label{fig:activation energy}
    \end{figure}

   \textit{Temperature dependence of resistivity}:
    The temperature-dependent resistance of our devices~(Fig.~4\textbf{a} in the main text) displays a pronounced foot at low resistance values, while the apparent transition temperature~[defined by the half-point of $R(T)$] seems to depend on the angle $\theta$ between the in-plane field and the current.  Arrhenius plots of these data reveals that the resistance is thermally activated in this regime, as it is expected from the nucleation of thermally excited phase slips~\cite{AmbegaokarHalperin1969,Tinkhambook}. As seen in Fig.~\ref{fig:activation energy}, the phase slips lead to a measurable resistance extending about 100 mK below $T_\mathrm{c}$. From the common onset of the resistance downturn, it is evident that it is not the transition temperature on the Al film that depends on $\theta$, but the activation energies $E_\mathrm{A}(\theta)$ extracted from the Arrhenius plots. The latter are plotted in the inset to Fig.~\ref{fig:activation energy}. The angle dependence nicely follows that of the Josephson-coupling energy $E_\mathrm{J}$. For a sinusoidal current--phase relation~(CPR), the activation energy is given by $E_\mathrm{A}=2E_\mathrm{J}=2\Phi_0I_0/2\pi=2(\Phi_0/2\pi)^2/L_0$.
   Hence, we can estimate $2E_\mathrm{J}$ from the first expansion coefficient $L_0(\theta)$ of the Taylor expansion of $L(I,\theta)$ with respect to $I$ (see Tab.~\ref{tab:activation energy} and Fig.~2\textbf{b} in the main text), if we make the simplifying assumption of a sinusoidal shape of the CPR. This way, the values of $2E_\mathrm{J}$ are overestimated by a factor $\simeq2$, while the angle dependence agrees very nicely. The difference between $E_\mathrm{A}$ and $2E_\mathrm{J}$ may come from the strongly nonsinusoidal character of the CPR. At the lowest resistance values $R\lesssim200\,\Omega$, the slope of the Arrhenius plot seems to change. As a single junction contributes about $67\,\Omega$ to the total resistance, it seems that about three junctions display a reduced Josephson-coupling energy. It is these junctions that likely determine the transport critical current of the array, which is about two times smaller than the more reliable value extracted from the Josephson inductance. Nevertheless, the qualitative agreement between the two energy scales extracted from entirely independent experiments is remarkable.

\begin{center}
	\begin{table}[t]
	    \caption{Characteristic parameters of sample~2 (see Figs.~\ref{fig:other samples}, \ref{fig:activation energy}).}
		\begin{tabular}{| l | l | l | l | l |}
			\hline
			$\theta$  &$L_{\mathrm{0}}$ (pH)  & $I_{\mathrm{0}}$ ($\mu$A) & 2$E_{\mathrm{J}}$ (meV) & $E_{\mathrm{A}}$ (meV)\\ \hline
			0°   & 174.6 &  1.82 & 7.49 & 4.00\\ \hline
			15°   &  171.9 & 1.85 & 7.61 &   \\ \hline
			22.5°   &    &   &   & 4.02 \\ \hline
			30° &   169.5 & 1.88 & 7.72 & \\ \hline
			45°	&  166.1 & 1.92 & 7.87 & 4.31\\ \hline
			60° & 161.3 & 1.97 & 8.10 & \\ \hline
			67.5°	&   &  & &  4.66\\ \hline
		    75°  &  160.5 & 1.98 & 8.15 & \\ \hline
			90° &  158.7 & 2.00 & 8.23 & 4.75\\ \hline
		\end{tabular}
		 \label{tab:activation energy}
	\end{table}
\end{center}

\begin{figure}[t]
        \centering
        \includegraphics[width=0.475\textwidth]{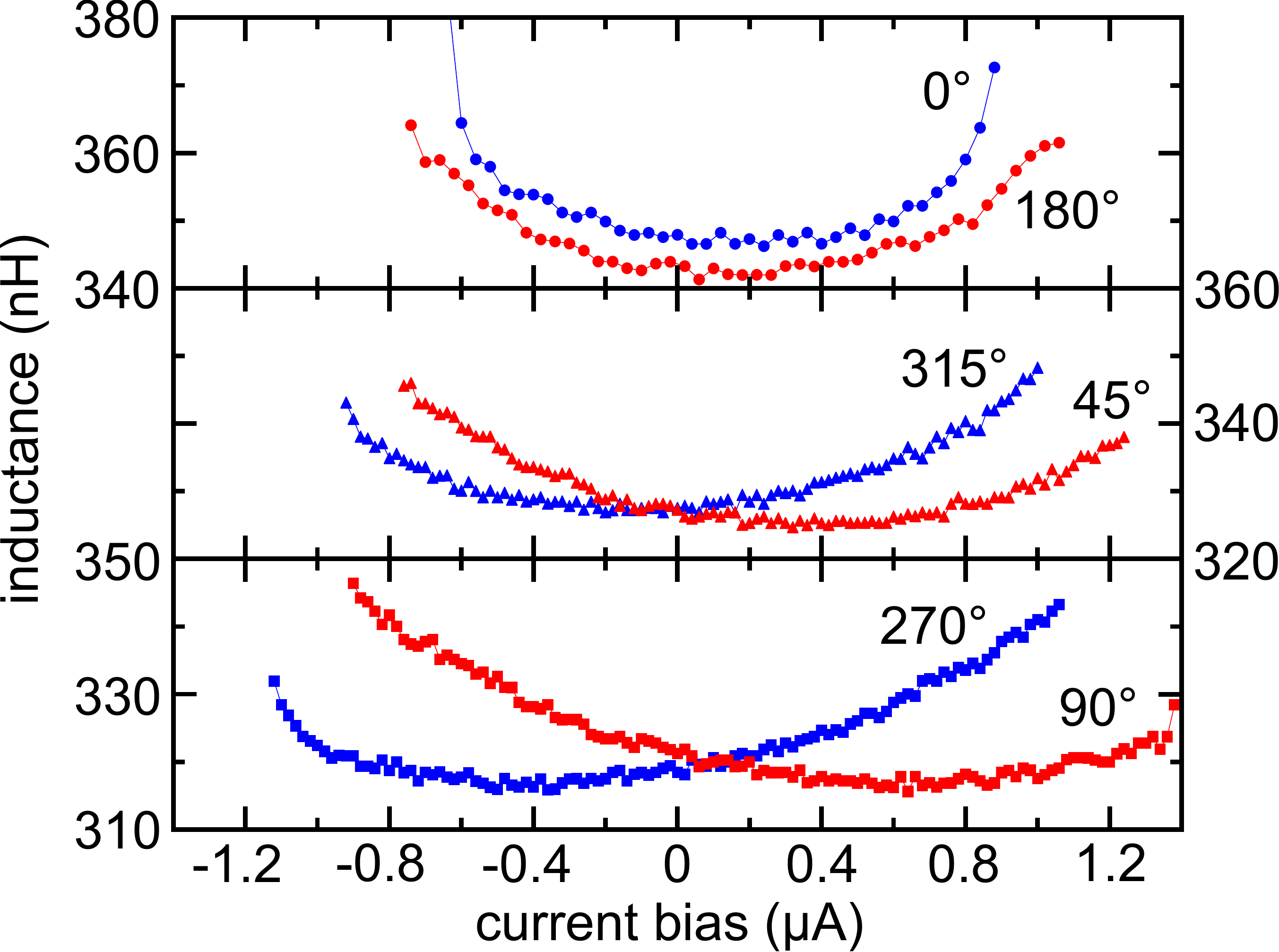}
        \caption{Data in Fig.~2\textbf{a} of the main text, plotted without applying offsets.
        }
        \label{fig:offs}
    \end{figure}

    \textit{Offset in Fig.~2\textbf{a} of the main text: } In the main text (caption of Fig.~2) we have mentioned that a small vertical offset was applied to the $L(I)$ data. Here, we comment more about this offset and its experimental origin. Data without offset is shown in Fig.~\ref{fig:offs}.
    The offset is caused by a small residual field due to trapped vortices in the superconducting coil. 
		Such a residual field is typically of the order of a few mT. It renders an asymmetry of the applied in-plane field: as an example, for nominally applied $\pm 100$~mT one obtains instead +98/-102~mT. As a consequence of this asymmetry, we observe a  small change of the minimum inductance $L_{min}$ (related to the change of critical current), of the order of 2-3 nH (roughly 1\%). This is, however, visible in  Fig.~2\textbf{a} owing to the strong magnification of the vertical axis. 
		On the other hand, the impact of this offset on the anomalous shifts ($\varphi_0$, $ \varphi^{\ast}$, $i^{\ast}$, etc.) is minimal. 
		For better readability of Fig.~2\textbf{a} we decided to remove this spurious offset in the main text, by vertically matching the minima of $L(I)$ for the two field polarities (red and blue curves).
    
    Finally, we have subtracted a constant offset of +130~nA in the current axis from all $L(I)$ curves. Such offset is trivially due to voltage offsets between the current-source instrument and the cold ground.  In later measurements of samples 2 and 3, we have avoided the residual field by demagnetizing the magnet coil. For these samples, we have also \textit{measured} the DC current offset directly.

    \section{Spin-orbit coupling parameters from self-consistent $\mathbf{k} \cdot \mathbf{p}$~calculations}

\begin{figure*}[ht]
        \centering
        \includegraphics[width=\textwidth]{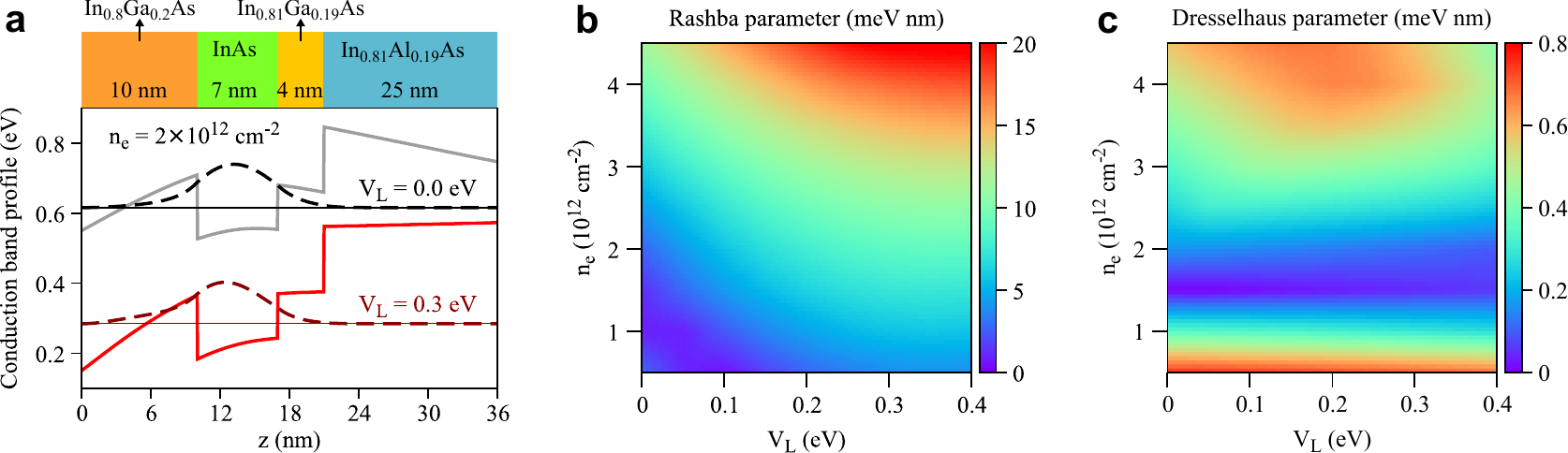}
        \caption{\textbf{a},~Conduction-band profiles of the studied InAs-based quantum well for two representative values of $ V_\mathrm{L} $---defining the boundary condition at the left interface (one close to Al layer)---and an electron density of $ n_\mathrm{e} = 2 \cdot 10^{12} \; \textrm{cm}^{-2} $. The energy of the lowest subband is given by the horizontal thin line and its probability density is shown with dashed lines. The different regions of the heterostructure sample, as well as their alloy composition, are shown on top of the plot. \textbf{b},~Rashba, $\alpha$, and \textbf{c},~Dresselhaus, $\beta_1 + \beta_{1,3}$, parameters as functions of $ V_\mathrm{L} $ and $ n_\mathrm{e} $.}
        \label{Fig_kp_soc}
    \end{figure*}

    In order to estimate realistic values of the spin-orbit coupling parameters for the InAs-based heterostructure, we investigate the electronic structure of the conduction-band by solving self-consistently the Schr\"{o}dinger--Poisson equation
    within the Hartree approximation~\cite{Calsaverini2008PRB,Fu2015PRB,Antipov2018PRX,Mikkelsen2018PRX}. We first solve the problem self-consistently in the absence of spin-orbit coupling and then analyze the two most important spin-orbit coupling contributions that can arise in a two-dimensional electron gas: the Bychkov--Rashba term~\cite{Bychkov1984,Bychkov1984b,*Bychkov1984c}, due to the inversion asymmetry of the heterostructure, and the Dresselhaus term~\cite{Dresselhaus1955}, due to the bulk inversion asymmetry of the constituent materials. The conduction electrons confined in the InAs quantum well are subject to the effective spin-orbit coupling Hamiltonian~\cite{Koralek2009Nat,Fu2015PRB}
    \begin{align}
    \hat{\mathcal{H}}_{\text{SOC}} & =\alpha\left(k_{x}\sigma_{y}-k_{y}\sigma_{x}\right)\nonumber \\
     & +\beta_{1}\left(k_{x}\sigma_{x}-k_{y}\sigma_{y}\right)\nonumber \\
     & +\frac{4\beta_{1,3}}{k_\mathrm{F}^{2}}\left(k_{x}k_{y}^{2}\sigma_{x}-k_{y}k_{x}^{2}\sigma_{y}\right) \, ,
    \end{align}
    in which $\alpha=\left\langle \alpha(z)\right\rangle$ is the Rashba parameter, $\beta_{1}=\left\langle \partial_{z}\gamma(z)\partial_{z}\right\rangle$ is the linear Dresselhaus parameter, $\beta_{1,3}=\left\langle \gamma(z)\right\rangle k_\mathrm{F}^{2}/4$ is the linearized cubic Dresselhaus parameter, and $k_\mathrm{F}^{2}=2m_{s}^{*}\left(E_\mathrm{F}-E_{s}\right)/\hbar^{2}$ is the  square of the Fermi wave vector of the electron subband with effective mass $m_{s}^{*}$ and energy $E_{s}$. The angle brackets indicate the averaging~$\left\langle o(z)\right\rangle = \int_L \text{d} z \, f^*(z) o(z) f(z)$, in which $L$ is the length of the system along the $\hat{z}$-direction and $f(z)$ is the envelope function of the corresponding electron subband.  
    The profile of the Rashba coupling $\alpha(z)$ is obtained from 
    the conventional $ 8 \times 8 $ Kane~model~\cite{Kane1957,Winkler2003spin} in terms of 
    $\mathbf{k} \cdot \mathbf{p}$ downfolding. 
    We follow the procedure of Refs.~\cite{Calsaverini2008PRB,Fu2015PRB}, but take the full spatial dependence of the parameters in the Kane~model into account, thus obtaining 
    \begin{align}
    \alpha(z) & =\frac{P(z)P^{\prime}(z)}{3}\left[E_{1}(z)+E_{2}V(z)\right]\nonumber \\
     & +\frac{P^{2}(z)}{3}\left[E_{1}^{\prime}(z)+E_{2}^{\prime}(z)V(z)+E_{2}(z)V^{\prime}(z)\right] \, ,
    \end{align}
    in which $V(z)$ contains the band~offsets and electrostatic potentials, and
    \begin{align}
    E_{1}(z) & =\frac{1}{E_\mathrm{g}(z)+\Delta_\mathrm{SO}(z)}-\frac{1}{E_\mathrm{g}(z)}\nonumber \\
    E_{2}(z)= & \frac{1}{\left[E_\mathrm{g}(z)+\Delta_\mathrm{SO}(z)\right]^{2}}-\frac{1}{\left[E_\mathrm{g}(z)\right]^{2}} \, ;
    \end{align}
    $ E_\mathrm{g}(z) $ corresponds to the spatially dependent bandgap and $ \Delta_\mathrm{SO}(z) $ to the spin-orbit splitting. 
    For the Poisson~equation, we employ the Dirichlet boundary~conditions by setting the electrostatic potential to $-V_\mathrm{L}$ for the left side (at the interface with the Al~layer) and to zero for the opposite side of the system (see Fig.~\ref{Fig_kp_soc}\textbf{a} for the schematic view of the heterostructure region considered in the calculations). Physically, this boundary condition introduces a linear electric field across the heterostructure, simulating the influence of an external gating~\cite{Calsaverini2008PRB,Fu2015PRB} or mimicking band-bending effects due to charge transfer to/from the Al layer~\cite{Antipov2018PRX,Mikkelsen2018PRX}. It has been shown that the use of such linear electric fields provides reliable estimates of the spin-orbit coupling parameters in nanostructures~\cite{Calsaverini2008PRB,Fu2015PRB,Campos2018PRB,Mayer2020}. The material parameters used in the calculations are taken from Ref.~\cite{Vurgaftman2001}, except for the bulk Dresselhaus parameters, which can be found in~Ref.~\cite{Chantis2006PRL}.
    
    Our calculations for the InAs quantum well system are summarized in Fig.~\ref{Fig_kp_soc}. In Fig.~\ref{Fig_kp_soc}\textbf{a}, we show the region of the heterostructure considered for the calculations and the conduction-band profile obtained self-consistently for two representative values of $ V_\mathrm{L} $~(0 and 0.3 eV) and an electron density of $ n_\mathrm{e} = 2 \cdot 10^{12} \, \textrm{cm}^{-2} $. The energy of the first electron subband, as well as its probability density, are also presented. The Rashba, $\alpha$, and Dresselhaus, $\beta_1 + \beta_{1,3}$, parameters are shown in Figs.~\ref{Fig_kp_soc}\textbf{b}--\textbf{c}, respectively, as functions of $ V_\mathrm{L} $ and $ n_\mathrm{e} $. For values of $ V_\mathrm{L} \gtrsim 0.1$ eV and $ n_\mathrm{e} \gtrsim 1.5 \cdot 10^{12} \; \textrm{cm}^{-2}$, the Rashba parameter exceeds the value of the Dresselhaus and is the dominant spin-orbit coupling parameter of the system. 
    
    \section{\textsc{Kwant} transport simulations}


    \subsection{Theoretical model}
    
    We assume a three-dimensional InAs quantum~well spreading to infinity along the longitudinal $ \hat{x} \parallel [110] $~direction, while its dimensions along the transverse $ \hat{y} \parallel [1 \overline{1} 0] $ and $ \hat{z} \parallel [001] $~directions are finite, and given by the width~$ w $ and height~$ h $, respectively. 
    The coordinate~system~\footnote{As a comment, in the $\mathbf{k} \cdot \mathbf{p}$ study, the $\hat{z}$-axis has shifted origin and reversed orientation. For the~\textsc{Kwant}~implementation, it is more convenient to use the present coordinate system. However, it is clear that the two different conventions lead to the same physical conclusions.} is chosen such that the quantum~well covers the transverse real-space~regions~$ y \in [-w/2;w/2] $ and $ z \in [-h/2;h/2] $. 
    To implement \emph{a single Josephson~junction}, we model the Al/InAs~heterostructure as two semi-infinite superconductors~(with  induced superconducting gap $\Delta^*$) that are separated by a weak link consisting of a short 
    nonsuperconducting region within the interval~$ 0 \leq x \leq d $. 
    The spectral features of such junctions are captured by the Bogoljubov--de~Gennes~Hamiltonian~\cite{DeGennes1989}
    \begin{equation}
        \hat{\mathcal{H}}_\mathrm{BdG} = \left[ \begin{matrix} \hat{\mathcal{H}}_\mathrm{e} & \hat{\Delta} \\ \hat{\Delta}^\dagger & \hat{\mathcal{H}}_\mathrm{h} \end{matrix} \right] ,
        \label{EqBdG}
    \end{equation}
    where the single-electron Hamiltonian reads as
    \allowdisplaybreaks
    \begin{align}
        \hat{\mathcal{H}}_\mathrm{e} &= \left[ -\frac{\hbar^2}{2m^*} \left( \frac{\partial^2}{\partial x^2} + \frac{\partial^2}{\partial y^2} + \frac{\partial^2}{\partial z^2} \right) - \mu \right] \hat{\sigma}_0 \nonumber \\[5 pt]
        &\hspace{25 pt} + V_\mathrm{B} \hat{\sigma}_0 \Theta(x) \Theta(-x + d_\mathrm{B}) \nonumber \\[5 pt]
        &\hspace{25 pt} + V_\mathrm{B} \hat{\sigma}_0  \Theta(x - d_\mathrm{B} - d) \Theta(-x + 2 d_\mathrm{B} + d) \nonumber \\[5 pt]
        &\hspace{25 pt} + \alpha (k_y \hat{\sigma}_x - k_x \hat{\sigma}_y) \nonumber \\[5 pt]
        &\hspace{25 pt} + \frac{1}{2} g^* \mu_\mathrm{B} B_\mathrm{ip} (\hat{\mathbf{m}} \cdot \hat{\boldsymbol{\sigma}}) \nonumber \\[5 pt]
        &\hspace{25 pt} + \frac{1}{2} m^* \Omega^2 z^2 \hat{\sigma}_0 ,
        \label{EqSingleElectronHam}
    \end{align}
    and its hole counterpart is given by~$ \hat{\mathcal{H}}_\mathrm{h} = -\hat{\sigma}_y \hat{\mathcal{H}}_\mathrm{e}^* \hat{\sigma_y} $; in the above expressions, $ \hat{\sigma}_0 $ denotes the two-by-two identity matrix
    and $ \hat{\sigma}_i $ the $ i $th Pauli~spin~matrix. 
    
    The first line in $\hat{\mathcal{H}}_\mathrm{e}$ gives the kinetic energy of electrons with the effective mass~$ m^* \approx 0.02 m_0 $~($ m_0 $ is the free-electron~mass)~\cite{Vurgaftman2001,Fabian2007}, measured from the Fermi~level of the \emph{uncovered} InAs~2DEG,~$ \mu = (\hbar^2 k_\mathrm{F}^2) / (2m^*) \approx 239 \, \mathrm{meV} $. 
    This value corresponds to an electron~density of~$ n_\mathrm{e} \approx 2 \cdot 10^{12}\, \mathrm{cm}^{-2} $, as indicated by  Hall-contact measurements.
    
    To model a slightly reduced junction transparency, we insert a weak potential~barrier with an energy height~$ V_\mathrm{B} $ and spatial width~$ d_\mathrm{B} \ll d $ between the superconducting regions and the central nonsuperconducting link---see the second and third line in Eq.~\ref{EqSingleElectronHam}. 
    
    The broken inversion~symmetry along the sample growth direction gives rise to strong Rashba spin-orbit coupling inside the quantum~well---the fourth line in Eq.~\ref{EqSingleElectronHam}. Its strength~$ \alpha $ generally decreases with increasing well thickness~$ h $. 
    An earlier work~\cite{Mayer2020} estimated $ \alpha \approx 15 \, \mathrm{meV} \, \mathrm{nm} $ for $ h = 10 \, \mathrm{nm} $, 
    while a slightly larger value of $ \alpha \approx 28 \, \mathrm{meV} \, \mathrm{nm} $ was reported in~Ref.~\cite{Shabani2016}.
    A similar range of $\alpha\approx$~$10 $--$ 15 \, \mathrm{meV} \, \mathrm{nm} $ is also obtained from our self-consistent $ \mathbf{k} \cdot \mathbf{p} $ calculations, which take the experimentally estimated electron~density of~$ n_\mathrm{e} \gtrsim 2 \cdot 10^{12} \, \mathrm{cm}^{-2} $, sample height $ h^\mathrm{exp.} = 7 \, \mathrm{nm} $, and electrostatic gating into account. 
    Following our $ \mathbf{k} \cdot \mathbf{p} $~analysis, the Dresselhaus contribution is expected to be about one order of magnitude smaller and is therefore not included into our model Hamiltonian. 
    On the contrary, the
    Rashba spin-orbit~coupling enters as the conventional term \emph{linear}~in momentum. 
    As a comment, a recent study~\cite{Alidoust2021} demonstrated that if pure \emph{cubic} spin-orbit~coupling~terms would become dominant, the shape of the current--phase~relations would alter. However, our experimental findings do not show any of such features.

    Additionally, the time-reversal~symmetry is broken~(apart from the passing supercurrent) by an in-plane~($ xy $-plane) magnetic~field that is parameterized by its magnitude~$ B_\mathrm{ip} $ and direction~vector~$ \hat{\mathbf{m}} = [ \cos \theta, \, \sin \theta, \, 0 ]^\top $; the angle $\theta$ is measured with respect to the current-flow $ \hat{x} $~direction. 
    We assume that the magnetic field couples just with the spin degrees of freedom and ignore orbital effects---see the fifth line in Eq.~\ref{EqSingleElectronHam}, where $ \mu_\mathrm{B} $ stands for the Bohr~magneton and $g^*$ for the in-plane g-factor of the InAs quantum~well. 
    The latter was estimated as~$ g^* \approx -10 $ in Ref.~\cite{Mayer2020}. 
    Generally, the $ g $-factors of quantum~wells are typically smaller than the corresponding bulk values, for example,~$ g^*_\mathrm{bulk} \approx -15 $ for InAs~\cite{Mayer2020}.

\begin{figure}[t]
        \centering
        \includegraphics[width=0.50\textwidth]{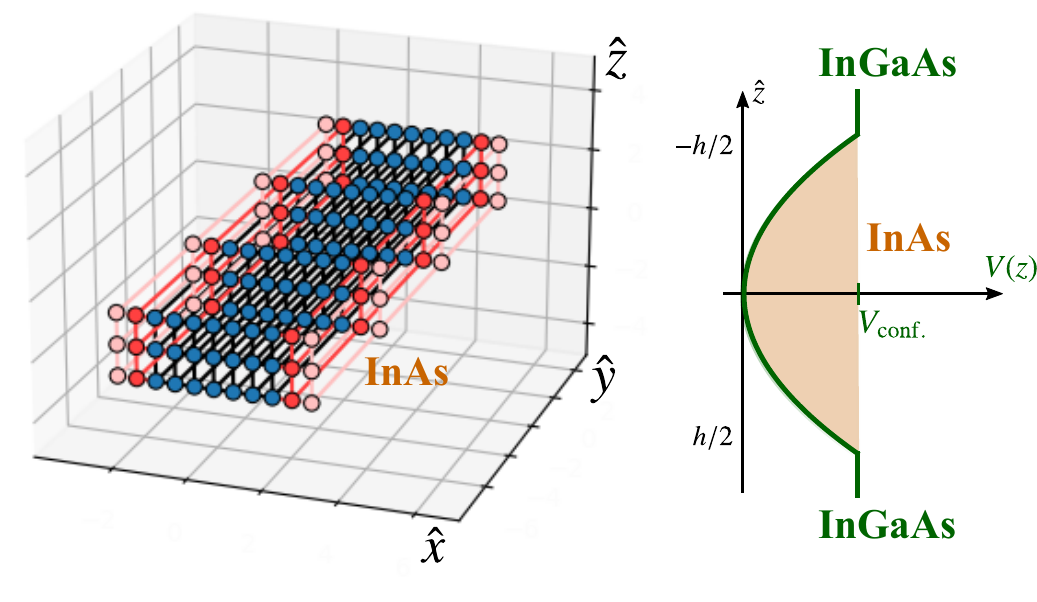}
        \caption{\textsc{Left:}~Schematics of the generic three-dimensional tight-binding~grid that discretizes a \emph{single InAs~quantum-well Josephson~junction} with respect to the crystallographic directions~$ \hat{x} \parallel [110] $, $ \hat{y} \parallel [1 \overline{1} 0] $, and~$ \hat{z} \parallel [001] $. 
        The semi-infinite superconducting~electrodes~(red) span the regions~$ x \leq 0 $ and $ x \geq d $~(neglecting, for simplicity, the interfacial barriers), whereas the nonsuperconducting weak link (blue) occupies~$ 0 \leq x \leq d $; the quantum~well has the width $ -w/2 \leq y \leq w/2 $ and height~$ -h/2 \leq z \leq h/2 $. 
        The colored dots indicate the on-site~energies resulting from discretizing the Bogoljubov--de~Gennes~Hamiltonian in~Eq.~\ref{EqBdG}, whereas the black and colored lines highlight the corresponding hopping~terms. 
        \textsc{Right:}~Schematic sketch of the parabolic confinement~potential in InAs~well that confines electrons in the vertical ($\hat{z}$) direction with the maximal amplitude~$ V_\mathrm{conf} $; in real samples, this confinement would result from the conduction-band~mismatches between the InAs and neighboring InGaAs~layers. 
            }
        \label{Fig_Scheme}
    \end{figure}

    The band offset between the InAs~quantum~well and the adjacent In$ _{0.8} $Ga$ _{0.2} $As and In$ _{0.81} $Ga$ _{0.19} $As layers results in a material-specific profile of the electrostatic potential along the $ \hat{z} $-direction---see Fig.~\ref{Fig_kp_soc}\textbf{a} displaying the self-consistently computed conduction-band profile for two representative values of~$ V_\mathrm{L} $. 
    In order to model such a profile in our transport~simulations---see the last line in Eq.~\ref{EqSingleElectronHam}---we approximate it inside the InAs~well by an effective parabolic confinement~$ V(z) \equiv m^* \Omega^2 z^2 \hat{\sigma}_0 / 2 $, where 
    $ \Omega $ abbreviates~$ \Omega = 2 \sqrt{2 V_\mathrm{conf} / m^*} / h $; see~Fig.~\ref{Fig_Scheme} for a schematic sketch.  
    The value of $V_\mathrm{conf}$ can be estimated as follows.
    The conduction-band~offsets in an Al$ _X $Ga$ _{1-X} $/GaAs/Al$ _X $Ga$ _{1-X} $~quantum~well~\cite{Seraide2002} causes a confinement potential with the magnitude~\footnote{Note that we corrected a typo in~$ V_\mathrm{conf}(X) $ that has been present in~Ref.~\cite{Seraide2002}.} $ V_\mathrm{conf}(X) = 0.6 \cdot (1.155X + 0.37X^2) \, \mathrm{eV} $, which suggests $ V_\mathrm{conf} \approx 150 \, \mathrm{meV} $ at the~stochiometric ratio $ X=0.2 $.
    Let us stress that the precise shape of the confinement potential is not so essential for the forthcoming discussions and other potential~profiles, like the rectangular one, would provide the same qualitative results. For completeness, we emphasize that we simply assume hard-wall boundary~conditions along the $ \hat{y} $-direction since $ w \gg h $.

    The off-diagonal parts of the Bogoljubov--de~Gennes Hamiltonian~(Eq.~\ref{EqBdG})  couple electrons and holes through the proximity-induced superconducting pairing~potential
    \begin{equation}
        \hat{\Delta} = \Delta^* (T) \left[ \Theta(-x) + \mathrm{e}^{\mathrm{i} \varphi} \Theta(x - 2 d_\mathrm{B} - d ) \right] \hat{\sigma}_0 ,
    \end{equation}
    whose amplitude scales with temperature~$ T $ according to the conventional BCS formula 
    \begin{equation}
        \Delta^* (T) = 0.134 \, \mathrm{meV} \cdot \tanh \left( 1.74 \cdot \sqrt{\frac{1.33 \, \mathrm{K}}{T} - 1} \right) .
    \end{equation}
    As obvious, $ \varphi $ refers to the phase~difference between the two superconducting elements forming the Josephson junction. 
    Most of the experimental studies have been performed at~$ T = 100 \, \mathrm{mK} $, so we use the same value also in our theoretical modeling.

    To access the transport characteristics of the Josephson~junction---current-phase~relation, 
    Josephson inductance, etc.---we implement the Bogoljubov--de~Gennes~Hamiltonian $ \mathcal{\hat{H}}_\mathrm{BdG} $, see~Eq.~\ref{EqBdG}, into the Python-based~transport~package~\textsc{Kwant}~\cite{Groth2014}. 
    The Hamiltonian is discretized on a generic three-dimensional~grid spanned by the 
    lattice~vectors~$ \mathbf{x} = [a_x, \, 0, \, 0]^\top $, 
    $ \mathbf{y} = [0, \, a_y, \, 0]^\top $, and 
    $ \mathbf{z} = [0, \, 0, \, a_z]^\top $. 
    The lattice~spacings~$ a_x $, $ a_y $, and $ a_z $ need to be much smaller than the Fermi~wavelength to assure proper convergence of the results. 
    To meet a reasonable balance between reliable outcomes and manageable computation~times when dealing with rather large systems, we take~$ a_x = 0.5 \, \mathrm{nm} $, $ a_y = 5 \, \mathrm{nm} $, and~$ a_z = 1 \, \mathrm{nm} $. 
    The employed discretization~grid is shown for a small quantum~well in~Fig.~\ref{Fig_Scheme}.

    To proceed further, we follow the methodology outlined in Refs.~\cite{Ostroukh2016} and~\cite{Zuo2017}. The Josephson~current is usually evaluated in the normal weak-link region, where the electrical current is conserved. Adding a fictitious vertical lead therein, \textsc{Kwant}'s functionalities allow us to invert the discretized~Hamiltonian $ \check{\mathcal{H}}_\mathrm{BdG} $ to access the Green's~function~$ \check{\mathcal{G}} $ and to compute the electrical current~density~(and afterwards the Josephson~current~$ I \equiv I_x $) given by the expression~\cite{Ostroukh2016} 
    \small
    \begin{equation}
        \mathbf{I}(\mathbf{r}) = \frac{2 e k_\mathrm{B} T}{\hbar} \mathrm{Re} \left\{ \sum_{n} \mathrm{Tr} \left[ \langle \mathbf{r} | \check{\mathcal{G}}(\mathrm{i} \omega_n) | \mathbf{r} \rangle \left\langle \mathbf{r} \Bigg| \frac{\partial \check{\mathcal{H}}_\mathrm{BdG}}{\partial \mathbf{k}} \Bigg| \mathbf{r} \right\rangle \right] \right\} ,
        \label{EqCurrent}
    \end{equation}
    \normalsize
    where $ e $ denotes the positive elementary charge, $ k_\mathrm{B} $ is Boltzmann's constant, and $ \omega_n = (2n+1) \pi k_\mathrm{B} T $~(with integer~$ n $) refers to the fermionic Matsubara~frequencies~(in units of~$ 1/\hbar $).

    The computationally most demanding task while evaluating Eq.~(\ref{EqCurrent}) is the inversion of~$ \check{\mathcal{H}}_\mathrm{BdG} $, yielding the Green's~function~$ \check{\mathcal{G}} $. 
    Dealing with realistic quantum~wells, one usually gets a considerable number of grid-lattice~sites, and consequently, a large Hamiltonian matrix whose inversion becomes extremely time-consuming. 
    For that reasons, we limit our \textsc{Kwant}~simulations to a quantum~well with the transverse dimensions~$ w = 0.5 \, \upmu \mathrm{m} $ and~$ h = 8 \, \mathrm{nm} $, while the nonsuperconducting weak link extends over~$ d = 10 \, \mathrm{nm} $~(short-junction~limit). 
    These parameters ensure computational manageability and, simultaneously, they properly reproduce the multi-channel~features~\cite{Mayer2020b}, i.e., they give a large enough number of transverse modes~(along the $ \hat{y} $- and/or $ \hat{z} $-directions) that contribute to the Josephson-current. 
    The scaling~behavior on the model side gives~$ w / d = 500 \, \mathrm{nm} / 10 \, \mathrm{nm} = 50 \gg 1 $, while the corresponding experimental value reads as~$ w^\mathrm{exp.} / d^\mathrm{exp.} = 3150 \, \mathrm{nm} / 100 \, \mathrm{nm} = 31.5 \gg 1 $, implying that our numerical simulations are able to proportionally capture the experimental geometry.

    As already mentioned, to account for the reduced junction transparency, we insert an ultrathin potential barrier possessing thickness~$ d_\mathrm{B} = 5 \, \text{\AA} $ and height~$ V_\mathrm{B} = \mu $ for each superconductor--weak-link interface. 
    These barriers can be characterized through their associated Blonder--Tinkham--Klapwijk $ Z $-factor~\cite{Blonder1982}, for which we obtain~$ Z = (m^* V_\mathrm{B} d_\mathrm{B}) / (\hbar^2 k_\mathrm{F}) \approx 0.0885 $. Equivalently, one can use the effective normal-state barrier~transparency~$ \tau = 1 / (1+Z^2) \approx 99 \, \% $. 
    Our choice of $ d_\mathrm{B}$ and $ V_\mathrm{B}$ is motivated by our previous experimental study~\cite{baumgartner2020}, which yields an average transparency of the short junction of about~$ \tau \approx 94 \, \% $. 

    \subsection{Generic features of current--phase relations}

    From the microscopic point of view, Josephson~currents originate from the tunneling of individual Cooper~pairs through the nonsuperconducting link via Andreev~bound~states. 
    The lack of both inversion and time-reversal symmetry~(caused by Rashba spin-orbit~coupling and the in-plane magnetic~field) modifies the Andreev~spectrum by an additional $ \varphi_0 $-phase~shift 
    that imprints on the current--phase~relation $ I(\varphi + \varphi_0) $. 
    The latter becomes nonsinusoidal and \emph{substantially asymmetric}---apart from acquiring a $ \varphi_0 $-phase~shift~(i.e., a horizontal shift in the current--phase diagram), there also emerges an inflection at \emph{finite}~$\varphi^{\ast}$ and $i^{\ast}$, which finally causes $I(\varphi+\varphi_0) \neq -I(-\varphi+\varphi_0)$; see Fig.~1 of the main text for a qualitative illustration. 
    Since such distorted $ I(\varphi) $ remains $2\pi$-periodic in~$ \varphi $, it can be expanded in terms of sines and cosines. The presence of cosine~terms makes the positive- and negative-current branches asymmetric. 
    As a consequence, the maximal currents for positive and negative directions differ, giving rise to the so-called \emph{superconducting diode effect}~\cite{Ando2020}---observed and analyzed for the first time in the case of Josephson junctions in the present work. 
    Still, the anomalous shift~$ \varphi_0 $ remains a useful figure of merit of the asymmetry, as the relative magnitude of $ i^{\ast}/I_\mathrm{c} $~($ I_\mathrm{c} $ denotes the critical current) is of the order of 
    $ \varphi_0/\pi $. It is therefore interesting to know how $\phi_0$ scales with, say, the Rashba strength~$\alpha$ or the number of transverse modes~(hence, the strength of the confinement potential). 
    
    For one-dimensional Josephson~junctions with a single transverse channel, Buzdin concluded~\cite{Buzdin2008} 
    that~$ \varphi_0 = 4 \alpha d |g^*| \mu_\mathrm{B} B_\mathrm{ip} \sin \theta/(\hbar v_{\mathrm{F},x})^{2} $, where $ v_{\mathrm{F},x} $ denotes the $ \hat{x} $-component of the Fermi~velocity of the mode. 
    Substituting the parameters corresponding to our experiment, with~$ B_\mathrm{ip} = 100 \, \mathrm{mT} $, yields $ \varphi_0 \approx 0 $, meaning that there is effectively no measurable $ \varphi_0 $-shift within the single-channel transport~regime.

    Nevertheless, the situation becomes substantially different once the junction is in the multi-channel~regime with many transverse modes, i.e.~for $ w / d \gg 1 $. 
    Generalizing the above expression for~$ \varphi_0 $, we expect that the individual single-mode contributions add, although their magnitudes differ due to the different $ \hat{x} $-projected Fermi~velocities. 
    More precisely, the larger the transverse Fermi~momenta~$ k_{\mathrm{F},y} $ and $ k_{\mathrm{F},z} $ for a given mode at the Fermi level, the smaller is its $ \hat{x} $-components~$ k_{\mathrm{F},x} $,
    and because of the parabolic confinement, the corresponding $ v_{\mathrm{F},x} (\propto k_{\mathrm{F},x} )$ is notably reduced. 
    Since $\varphi_0 \propto v_{\mathrm{F},x}^{-2}$, the contributions from the transverse modes with large transverse momenta rise overproportionally. 
    In the overall current--phase~relation, we detect the weighted average of all individual channels and thus more sizable $ \varphi_0 $-shifts are expected~\cite{Mayer2020b} when compared to the single-channel~scenario.

\begin{figure}[tb]
        \centering
        \includegraphics[width=0.475\textwidth]{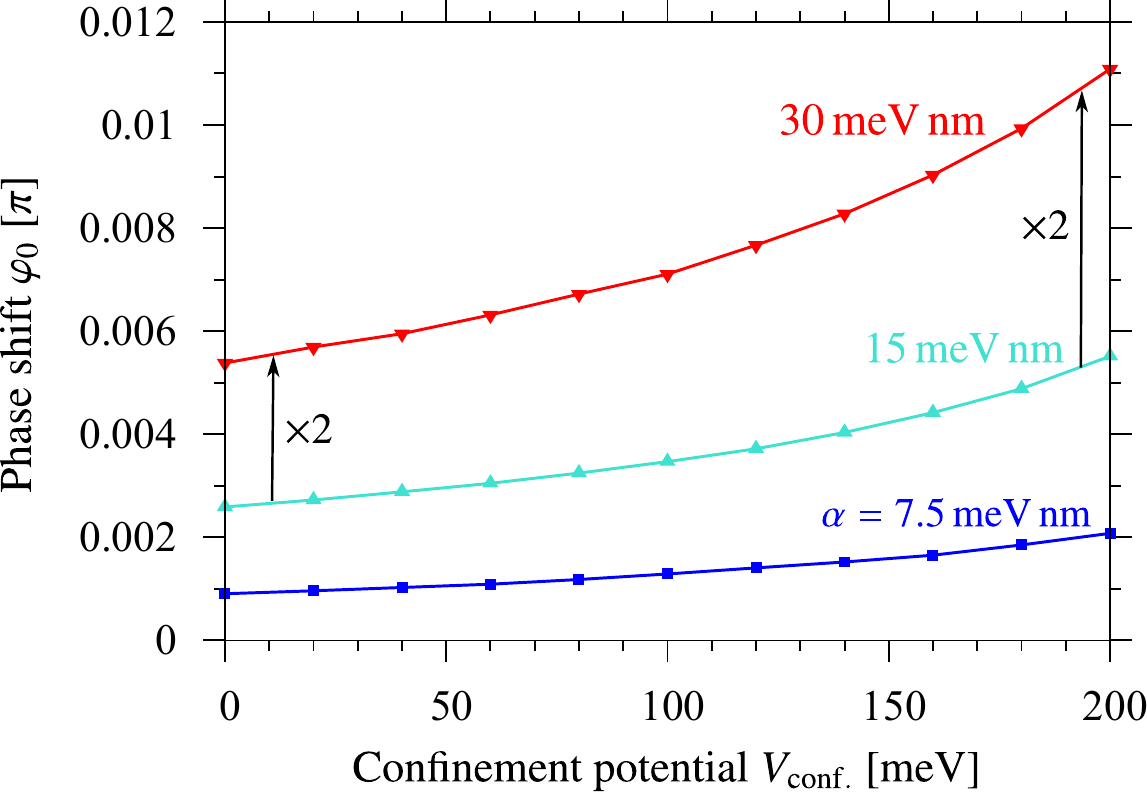}
        \caption{Calculated $ \varphi_0 $-shift as a function of the parabolic confinement~strength~$ V_\mathrm{conf} $ for a representative InAs quantum-well Josephson~junction with the spatial dimensions~$ w = 0.5 \, \upmu \mathrm{m} $, $ h = 8 \, \mathrm{nm} $, and $ d = 10 \, \mathrm{nm} $ to capture the multi-channel transport~regime, magnetic-field~strength~$ B_\mathrm{ip} = 100 \, \mathrm{mT} $, and angle~$ \theta = 90^\circ $. 
        }
        \label{Fig_Phi0}
    \end{figure}

    To quantitatively support our qualitative reasoning, Fig.~\ref{Fig_Phi0} illustrates the dependence of $ \varphi_0 $ on $ V_\mathrm{conf} $ for three representative Rashba spin-orbit~coupling strengths~$ \alpha = 7.5 \, \mathrm{meV} \, \mathrm{nm} $, $ \alpha = 15 \, \mathrm{meV} \, \mathrm{nm} $, and~$ \alpha = 30 \, \mathrm{meV} \, \mathrm{nm} $~(covering our estimates from the $ \mathbf{k} \cdot \mathbf{p} $~model). 
    Summarizing the results, $ \varphi_0 $ increases linearly with increased Rashba spin-orbit~coupling~$ \alpha $~(and with increased magnetic-field $ B_\mathrm{ip} $; not shown) in accordance with Buzdin's formula. 
    Furthermore, $ \varphi_0 $ increases nonlinearly~(weighted average of the individual transverse channels) as a function of the confinement potential~$ V_\mathrm{conf} $,  which has, to our best knowledge, not yet been discussed in earlier works. 
    In the simplest case, $ k_{\mathrm{F},y} = k_{\mathrm{F},z} \approx 0 $, and an enhancement of~$ V_\mathrm{conf} $ suppresses $ k_{\mathrm{F},x} $ according 
    to~$ k_{\mathrm{F},x} \propto \sqrt{1 - V_\mathrm{conf} / \mu} $, which becomes most relevant as $ V_\mathrm{conf} \to \mu $ with the corresponding substantial increase of~$ \varphi_0 $. 
    It is worth to mention that our numerical calculations resemble precisely this scaling of $ \varphi_0 $ with respect to~$ V_\mathrm{conf} $~(qualitatively and in good approximation also quantitatively, though nonzero $ k_{\mathrm{F},y} $ and $ k_{\mathrm{F},z} $ cause slight deviations from the simple formulas we stated).

    \subsection{Current--phase relation and Josephson inductance}

\begin{figure*}[tb]
        \centering
        \includegraphics[width=0.9\textwidth]{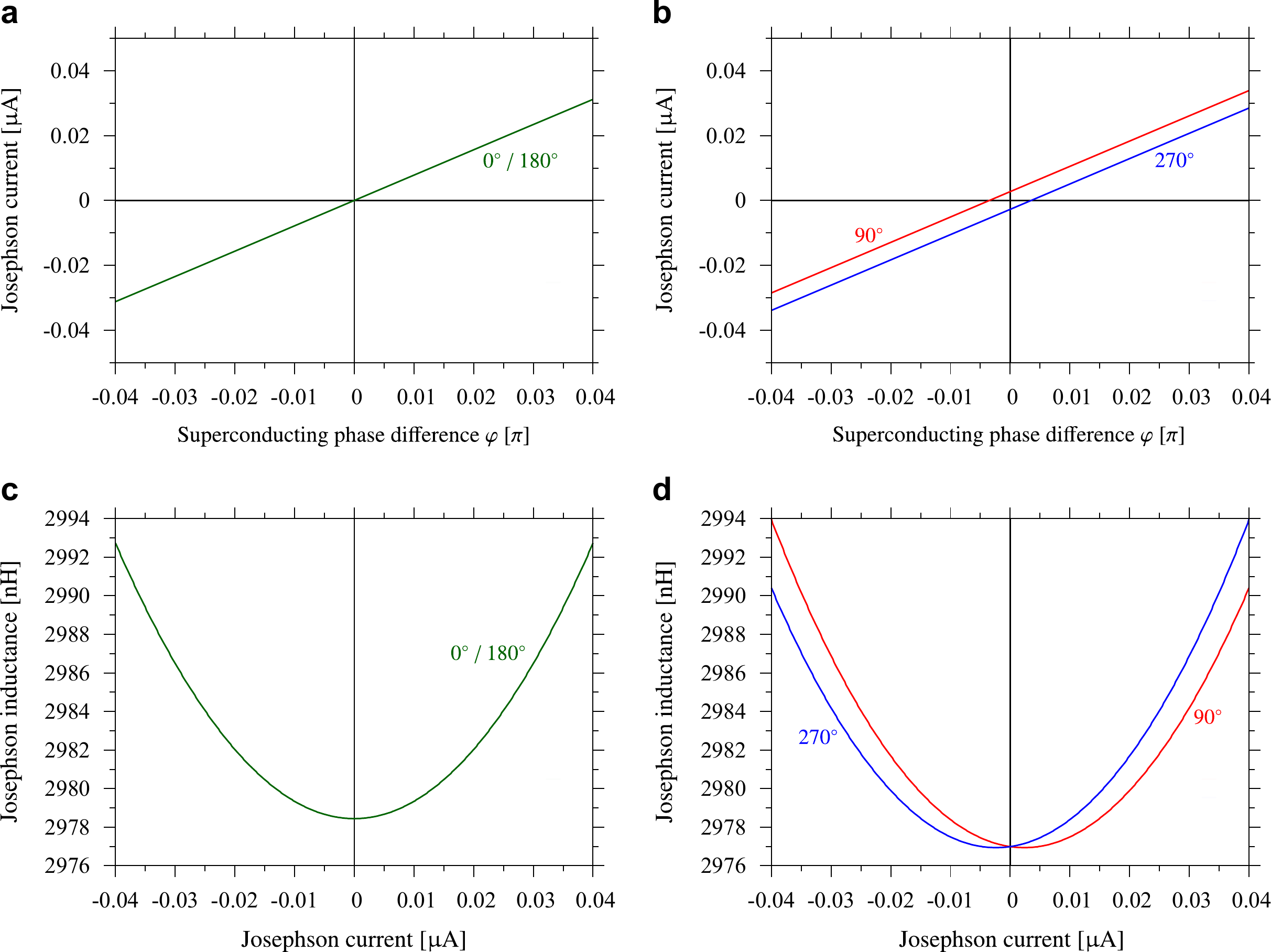}
        \caption{\textbf{a}, \textbf{b},~Calculated Josephson~current--phase~relations for a single InAs quantum-well Josephson~junction, assuming the spatial dimensions~$ w = 0.5 \, \upmu \mathrm{m} $, $ h = 8 \, \mathrm{nm} $, and $ d = 10 \, \mathrm{nm} $ to capture the multi-channel transport~regime, the representative parabolic potential~strength~$ V_\mathrm{conf} = 100 \, \mathrm{meV} $, as well as the magnetic-field~angles~(\textbf{a})~$ \theta = 0^\circ / 180^\circ $ and~(\textbf{b})~$ \theta = 90^\circ / 270^\circ $; the field's strength is~$ B_\mathrm{ip} = 100 \, \mathrm{mT} $. 
            \textbf{c}, \textbf{d},~The related Josephson~inductance--current~characteristics of the array consisting of~$ N = 2250 $ individual junctions in series, each with the above shown current--phase~relations. 
        }
        \label{Fig_Inductance}
    \end{figure*}

    The Josephson~inductance~$ L(\varphi) $ of the array of $ N = 2250 $ identical junctions is related to the current--phase~relation~$ I(\varphi) $ of an individual Josephson~junction via
    \begin{equation}
        L(\varphi) = \frac{\Phi_0}{2\pi} \cdot \left[ \frac{\partial I(\varphi)}{\partial \varphi} \right]^{-1} \cdot N ,
        \label{EqInductance}
    \end{equation}
    where~$ \Phi_0=h/(2e) $ denotes the magnetic flux~quantum. 
    In our calculations, we first compute the full current--phase~relation, from which we then extract all the quantities of interest like, e.g., the anomalous phase shift~$ \varphi_0 $, the two (different) critical currents~$ I_\mathrm{c}^+ $ and $ I_\mathrm{c}^- $, and the inflection~point~$ (i^{\ast},\varphi^{\ast}) $ of the current--phase~relation. In addition, we obtain the nonlinear inductance $L(I)$. Note that the minimum of $L(I)$ is located at the current $i^\ast$ at the inflection point. The position of this minimum is directly evident in our measurement of $L(I)$ and equivalent to the $\varphi_0$-shift, which cannot be determined by our experimental technique.
    In~Fig.~\ref{Fig_Inductance}, we present the Josephson~current--phase~relations together with the resulting $L(I)$-characteristics for the representative confinement~strength~$ V_\mathrm{conf} = 100 \, \mathrm{meV} $ and four different angles~$ \theta $ of the applied in-plane magnetic~field~(with respect to the $ \hat{x} \parallel [110] $-reference~direction).

    Let us first focus on the situation in which the magnetic~field is aligned (anti)parallel to the current~flow, i.e., either along~$ \hat{x} $~(meaning that~$ \theta = 0^\circ $) or along~$ -\hat{x} $~(indicating~$ \theta = 180^\circ $); see~Figs.~\ref{Fig_Inductance}\textbf{a} and~\textbf{c}. 
    Following our previous arguments, the $ \varphi_0 $-phase~shift in these two particular cases needs to vanish and the Josephson~current must approach its antisymmetric~current--phase~form,~$ I(-\varphi)=-I(\varphi) $, which turns out to be the sinusoidal $I(\varphi)\propto \sin \varphi \approx \varphi $ around zero phase~difference. Moreover, $ I(\varphi)$ is independent of the direction of the magnetic~field~(i.e., magnetic~fields along~$ \hat{x} $ and $ -\hat{x} $ yield equivalent results). 
    As a consequence, the current--phase~relation's inflection~point lies exactly at zero phase~difference and the corresponding $ L(I) $ becomes symmetric, with minimal Josephson inductance at zero current.

    A more interesting situation is displayed in~Figs.~\ref{Fig_Inductance}\textbf{b} and~\textbf{d}, where we consider the magnetic~field to point along~$ \hat{y} $~($ \theta = 90^\circ $) or~$ -\hat{y} $~($ \theta = 270^\circ $). 
    In these configurations, the interplay between the Rashba spin-orbit~coupling and the in-plane magnetic field induces nonzero $ \varphi_0 $-shifts in the current--phase~relations $I(\varphi+\varphi_0) $, with zero Josephson current at~$ - \varphi_0 $. 
    Since, according to Buzdin's formula, $ \varphi_0 \propto \sin \theta/v_{\mathrm{F},x}^2 $, the resulting phase~shifts are of the same magnitudes, but opposite signs. 
		The cosine terms in the CPR lead to \textit{finite} $i^{\ast}$, $\varphi^{\ast}$ and $\varphi_0$, resulting in asymmetric $ L(I) $-characteristics. 
    As expected, the calculated $ \varphi_0 $-shifts and the inductance-minima currents differ in signs, depending on the orientation of the in-plane field with respect to the reference~axis as seen in the experiment. This means that our theoretical model captures all necessary ingredients and nicely reproduces the experimental findings in a semi-quantitative way. 
    The calculations shown in~Fig.~\ref{Fig_Inductance} serve as motivation for the schematic illustrations provided in~Fig.~1 of the main~text.

    To quantify the inductance~asymmetries around zero current, we expand the~$ L(I) $-data up to second~order in~$ I $, $ L(I) \approx L_0 + L_0' I + L_0''/2 I^2 $, and extract the parameter 
    \begin{equation}
         -2 L_0' / (L_0 B_\mathrm{ip}) \ =\ \gamma_L\;\sin \theta\ ,
    \end{equation}
    where $\gamma_L$ is the newly defined supercurrent magnetochiral~anisotropy.
    While the absolute values of the inductance and of the Josephson~current depend on the number of transverse channels in the junctions (or, in other words, on its spatial extent), $ \gamma_L w$ constitutes a material-specific~(geometry-independent) quantity as soon as~$ w \gg h , d $. In that case, the finite-size~effects at the well's boundaries, which \textsc{Kwant} automatically includes into the calculation, do no longer play a substantial role. Therefore, the values obtained from our modeling are comparable to the experimental ones, even though the spatial dimensions are not the same.

    One of the most puzzling experimental observation is the surprisingly similar magnitude of the inductance magnetochiral~anisotropy deep inside the superconducting~state and the corresponding resistance magnetochiral~anisotropy close to the superconducting phase~transition. 
    If the amplitudes of the magnetochiral~anisotropies were predominantly controlled by the spin-orbit~coupling, one would expect a much larger value inside the superconducting~state, as it has been demonstrated for magnetoanisotropic transport in ferromagnet/superconductor~\cite{Hoegl2015,*Hoegl2015a} and superconductor/ferromagnet/superconductor~\cite{Costa2017} hybrid~structures. 
    This is usually associated with the energy~scale set by the superconducting~gap~$ \Delta^* $. 
    Owing to the gap's tiny values, the spin-orbit~coupling~energies~$ \varepsilon_\alpha = \alpha k_\mathrm{F} $ become sizable with respect to the reference~energy, $ \varepsilon_\alpha \gg \Delta^* $, and lead to dominant magnetoanisotropies.

    Regarding our work, the equal magnetochiral~anisotropies in both regimes~(well below and close to the critical~temperature) suggest that another ingredient must play an essential role---the (parabolic) confinement~potential along~$ \hat{z} $, as we illustrate in the main~text. 
    Apart from directly comparing the supercurrent magnetochiral anisotropy~amplitudes, there exists another confirmation that the magnetochiral~anisotropy results not only from the spin-orbit~coupling alone. 
    Inspecting the experimental data presented in~Fig.~2\textbf{b} of the main~text reveals that the constant inductance-expansion~coefficient~$ L_0 $ varies only marginally with the magnetic-field~angle, but nevertheless decreases with increasing~$ \theta $; i.e., the underlying current--phase~relation gets steeper around its inflection~point, which is probably related to a warping of the Fermi~surface in the parallel field. 
    This scaling can also only be theoretically reproduced within our simulations if we account for nonzero confinement~potential~$ V_\mathrm{conf} \gg 0 $.
    Setting~$ V_\mathrm{conf} = 0 $, $ L_0 $ would even increase with increasing~$ \theta $, which is in sharp contrast to the experimental outcomes.

\end{document}